\documentclass[pre,aps,10pt,superscriptaddress,notitlepage,nofootinbib]{revtex4-1}
\usepackage{amsmath}
\usepackage{amssymb}
\usepackage{enumerate}
\usepackage{graphicx}
\usepackage{color}

\newcommand{\refeq}[1]{(\ref{#1})}

\newenvironment{eqnn}{\begin{eqnarray}}{\end{eqnarray}}

\newcommand{\thav}[1]{\langle #1 \rangle}
\newcommand{\dav}[1]{\overline{#1}}
\newcommand{\s}{\sigma}
\newcommand{\dpart}[2]{\frac{\partial #1}{\partial{#2}}}
\newcommand{\dfunc}[2]{\frac{\delta #1}{\delta #2}}
\newcommand{\dtot}[2]{\frac{d#1}{d#2}}

\newcommand{\al}{\alpha}

\newcommand{\bx}{\mathbf{x}}
\newcommand{\by}{\mathbf{y}}
\newcommand{\bp}{\mathbf{p}}
\newcommand{\Sc}{\mathcal{S}}
\newcommand{\Z}{\mathcal{Z}}
\newcommand{\e}{\epsilon}
\newcommand{\de}{\delta}
\newcommand{\ox}{\overline{\bx}}
\newcommand{\oy}{\overline{\by}}
\renewcommand{\oc}{\overline{\chi}}
\newcommand{\bX}{\mathbf{X}}
\newcommand{\bY}{\mathbf{Y}}
\newcommand{\bA}{\mathbf{A}}
\newcommand{\br}{\mathbf{r}}
\newcommand{\hr}{\hat{\rho}}
\newcommand{\ux}{\underline{\bx}}
\newcommand{\uy}{\underline{\by}}
\newcommand{\erf}{\mathrm{erf}}

\begin{document}

\title{The replica method in liquid theory: from the basics to explicit computations}
\author{Corrado Rainone}
 \affiliation{LPT,
 Ecole Normale Sup\'erieure, CNRS UMR 8549, 24 Rue Lhomond, 75005 Paris, France}
\affiliation{Dipartimento di Fisica,
Sapienza Universit\`a di Roma,
P.le A. Moro 2, I-00185 Roma, Italy}

\maketitle

In this review we briefly introduce the fundamentals of the replica method in the context of liquid theory and the structural glass problem. In particular, we explain and show its usefulness as a computation framework in the context of the Random First Order Transition (RFOT) theory of the glass transition, whose defining points the reader is assumed to know. We shall give the intuitive idea of how and why the replica method is suitable for the description of the glass transition (the dynamical glass transition in particular) in real liquids, and then show how it can be used to make explicit computations and predictions that can be compared to experiments and numerical simulations.

\tableofcontents

\newpage

\section{Introduction}
The set of techniques that goes today under the name of ``replica method'' was first introduced in the context of schematic spin models for disordered alloys, like the Edwards-Anderson \cite{EA} (EA) or the Sherrington-Kirkpatrick (SK) \cite{SK} model, more commonly known as \emph{spin glasses}. In its first incarnation, it was called the ``replica trick'' and it was just a clever mathematical trick for computing the average over the probability distribution of the disorder that was necessary for the study of those models. Besides spin models, it also saw application in disordered models of pinned fluids, see for example \cite{menondasgupta,thalmanndasgupta}. At that stage, the method was just a comfortable computational tool and had no content form the point of view of the physics.\\
The idea that the replica trick was about something more than just computing a difficult average came about later, with the Parisi solution for the SK model \cite{parisiSG,parisiSK} and its subsequent physical interpretation \cite{parisiorderparameter,parisiSGandbeyond}. From that point onward, it started to become clear that introducing replicas of the system was not just part of a mathematical trick, but was also an elegant technique that allowed to probe the structure of the so-called free energy landscape, i.e. the structure of thermodynamical equilibrium states that the system can freeze in at low\footnote{We call ``low'' a temperature such that the phase space of the system becomes disconnected and ergodicity is broken.} temperature. Indeed, as we will show in the following, making a choice of a certain ``replica symmetry breaking'' (RSB) scheme is one and the same with formulating an hypothesis as to which this structure may be.\\ 
Some years later, the final ``emancipation'' of replicas from disorder took place with two works respectively from Monasson \cite{monasson} and Franz and Parisi \cite{franzparisi}, that showed how the usage of replicas has in principle nothing to do with disorder, but only with the structure of the free energy landscape of the system in study.

Indeed, the \emph{only} thing that all systems treated with the replica method (from spin glasses \cite{pedestrians} to structural glasses \cite{kirkpatrickthirumalaiwolynes} to constraint satisfaction problems \cite{mezardmontanari}) have in common is the fact that this landscape is, to put is simply, very \emph{rough}: the system has at his disposal lots and lots of minima which it can settle in at low temperature. This roughness has two main consequences: first of all, the presence of many minima has nontrivial effects on the entropy of the system; secondly, and differently from what happens with ordinary phase transitions, the pattern of symmetry breaking is not obvious: to each minimum corresponds a configuration of the local order parameter (the local magnetization for spin systems, the local density profile for particle systems) which has no visible symmetry or order, at least from our point of view. This inability to discern the pattern of symmetry breaking leads to another, and crucial, difficulty: 
we don'
t know what is the external field that we can use to select a state (i.e. to force the system to settle inside that particular state instead of all the others that it has at its disposal). When a state is associated with a magnetization which is just positive (or equivalently negative) at every point in space, as it happens with the two states in the Curie-Weiss model below the Curie temperature, it is trivial that an external field which is positive (negative) everywhere will select the positive (negative) state. If the magnetization has an effectively random behavior in space, however, not so much.\\
The replica method solves this problem by exploiting, in a nontrivial way, a fact which is however trivial in itself: we don't know what the right pinning field is, but the system does. It will simply be the configuration of the local magnetization (or the local density) in which the system freezes at low temperature. So the proposal is that we use the system itself as a pinning field: we make a copy, a \emph{replica} of it, thermalized at a certain temperature $T'$ (which is a priori different from the temperature $T$ of the original system), and then we use it to pin the original system, coupling them with a suitable parameter (called \emph{overlap} in the case of spin glasses) which measures the ``distance'' between the two. An alternative (and perhaps more intuitive) way of visualizing this is that even tough all the minima of the free energy landscape (or equivalently, all the possible configurations of the local order parameter) look the same to us because of disorder, they still look different to each other. So we probe 
the space of minima by comparing different replicas of the system. It is indeed easy to notice how any order parameter that one can come up with in a replica theory is about comparing configurations of the local order parameter between themselves, rather than with an ``ordered'' configuration that we are able to know \emph{a priori} as in the case of ordinary phase transitions.\\    
The field given by the replicated system will play the role of an external disorder. If we assume the distribution of the disorder to be uniform, then we get the Monasson ``real replica" method \cite{monasson}; if we assume it to be the canonical distribution at $T'$, we get the Franz-Parisi ``potential" method \cite{franzparisi}. There are some technical and practical differences between the two, but they lead to the exact same results, at least for what concerns equilibrium properties. We will focus here on the first one, as it is more straightforward and intuitive.

The fact that replicas have really to do with rough energy landscapes, and \emph{not really} with disordered averages, is often not stressed enough in pedagogical works. Since students who approach the subject are always presented (both for practical and historical reasons) first with the ``replica trick'' version the method and its application to spin glasses, and are taught only afterwards about its actual physical implications, some confusion can arise between the replica ``method'' and the replica ``trick''. This review presents the method in the context of glass-forming liquids (deterministic systems with no explicit disorder whatsoever) as way to further stress this point.\\
Another justification for this choice is the fact that, as of today, the most spectacular application of replicas comes about in the context of the glass transition problem. The amount of ongoing research in this sector is enormous and the ``glass community'' which produces it is one of the most active in condensed matter physics. Over the last few years in particular, replicas have allowed researchers to obtain results and quantitative predictions \cite{parisikurchan,zamponiurbani,fullRSB,zamponifull} which have shed new light both on the nature of the glass phase and on the physics of the jamming transition \cite{liunagel98,birolijamming}. From a method for the treatment of abstract spin models, replicas are now being used more and more as a powerful tool for problems in soft matter and materials science. This makes even more necessary an introduction of their fundamentals in a context which is near to those fields, and 
familiar to the people who practice them.\\
The theory that envisions, for structural glasses, a rough free energy landscape at low $T$ goes under the name of Random First Order Transition (RFOT) theory. As the glass problem is still far from a solution, it is still a point of (much heated) debate whether or not the RFOT scenario actually applies to glasses. As this review focuses on the replica method, we will assume the RFOT picture to be true, and the reader to be familiar with the physical content of it. We only want to show how and why replica theory is suitable for the study of systems with a rough free energy landscape, which we will presume as given. The interested reader can find excellent reviews on the subject of glasses and RFOT at references \cite{CavagnaLiq,berthierbiroli11}, and the original papers by Kirkpatrick, Thirumalai and Wolynes at references \cite{kirkpatrickthirumalai,kirkpatrickthirumalaiwolynes}.\\
Despite this omission of the physics of RFOT, the subject remains indeed vast and a detailed treatment of it would require at least a full monograph. This review's only ambition is to serve as an immediate and stimulating introduction to a huge end ever-evolving field. Our hope is that what we are going present here will stimulate the reader and enable him/her to approach the much more detailed and technical works on the field (see for example \cite{parisizamponi,jacquinthesis}) with interest and profit.

\section{The replica method\label{sec:1}}

\subsection{The problem}

\begin{figure}[hb!]
\begin{center}
\includegraphics[width=0.45\textwidth]{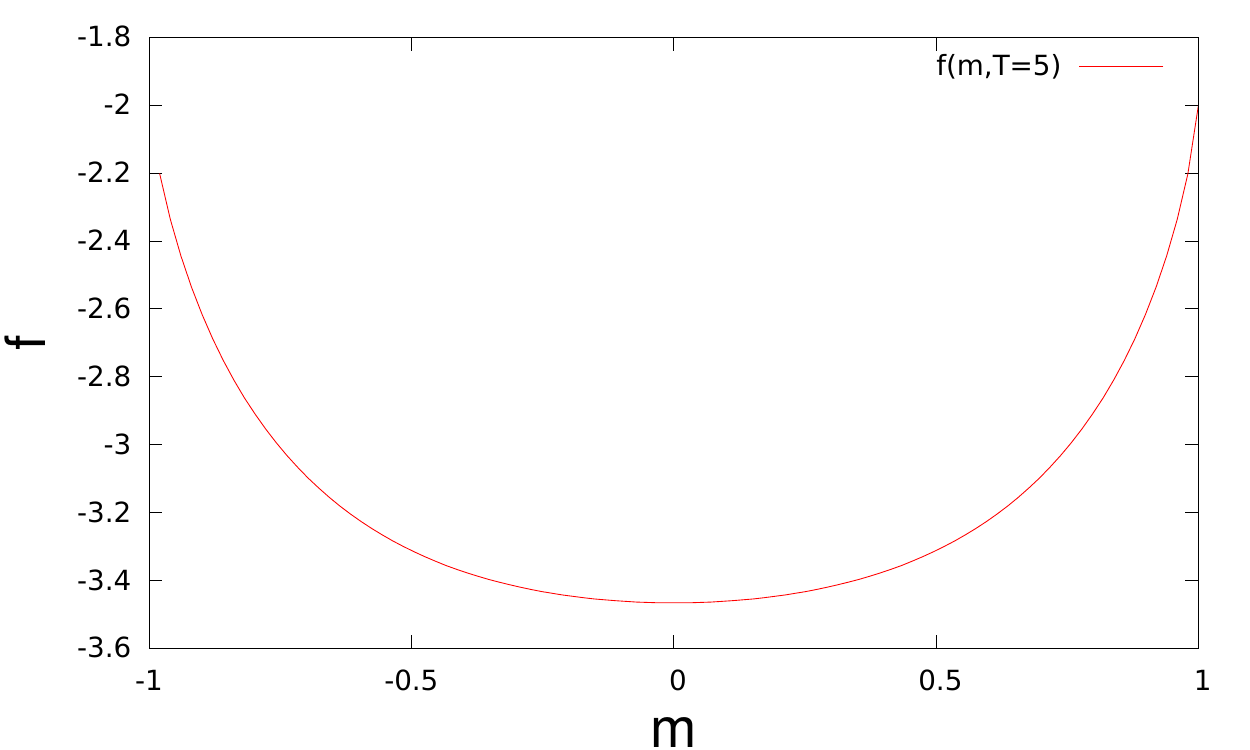}\nolinebreak[4]
\includegraphics[width=0.45\textwidth]{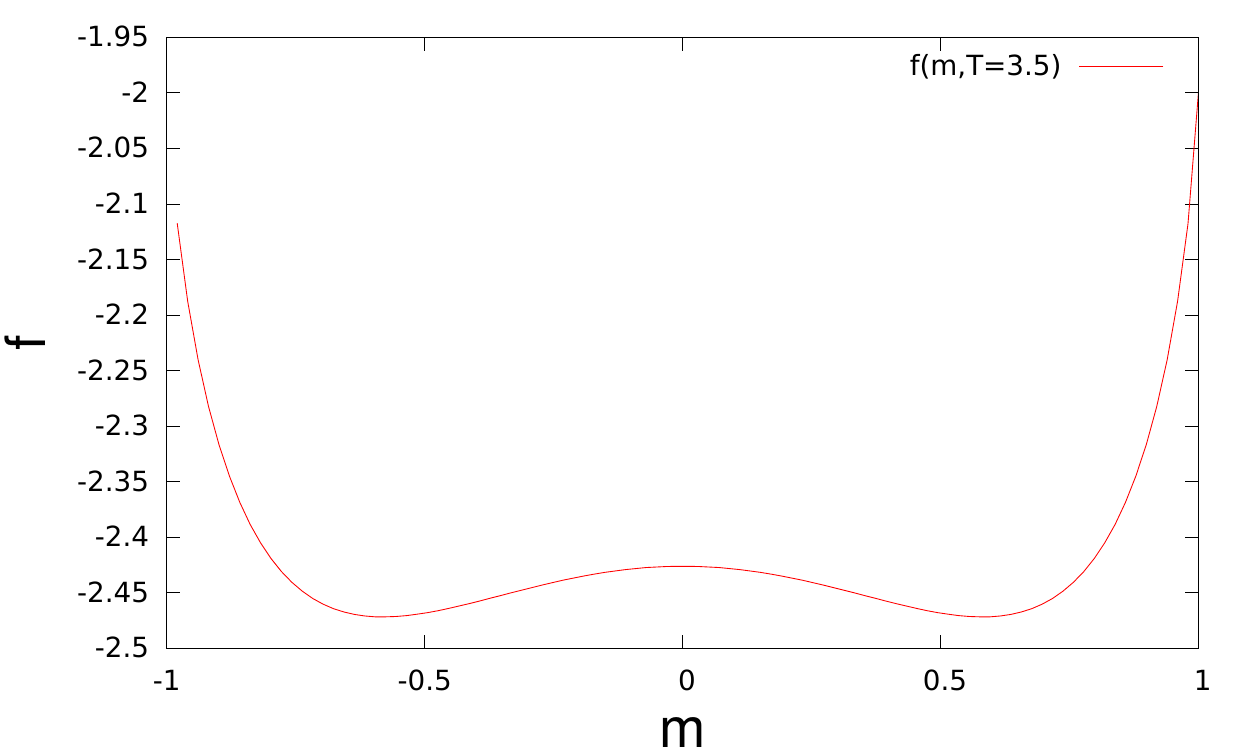}
\caption{Gibbs free energy as a function of the magnetization $m$ for the ferromagnetic Curie-Weiss model above (left) and below (right) $T_c$.}
\label{Ising}
\end{center}
\end{figure}

Let us consider a ferromagnetic Curie-Weiss model, whose Hamiltonian is
\begin{equation}
H = -\frac{1}{2N}\sum_{i,j}\sigma_i\sigma_j,
\label{Hising}
\end{equation}
where the sum extends over all pairs of spins, and the magnetization $m$ is defined as usual,
\begin{equation}
m = \frac{1}{N}\sum_{i=1}^N\thav{\sigma_i}.
\end{equation}
This model is the mean-field, fully connected variant of the Ising model. It is well known that under a certain temperature $T_c$, the model exhibits a phase transition from a paramagnetic, ergodic phase to a ferromagnetic phase wherein ergodicity is broken. The Gibbs free energy $f(m,\beta)$ for the model, as the system is cooled below $T_c$, is shown in figure \ref{Ising}.

We can see that above $T_c$ the free energy has a single paramagnetic minimum with zero magnetization. Below $T_c$, the free energy develops two degenerate ferromagnetic minima with opposite, nonzero magnetizations, which are linked by the spin-reversal symmetry of the Hamiltonian \refeq{Hising}
$$
\s \rightarrow -\s.
$$
Despite the fact that the two minima are linked by the symmetry, it is evident that once one of them has been chosen, the equilibrium state of the system is itself non-symmetric. This is the quintessential example of \emph{spontaneous symmetry breaking}.\\
Using a terminology taken from the theory of disordered systems, we will refer to those minima of the free energy as ``states''. Each state is labeled by the value of the magnetization in the corresponding minimum, which is given by stationarity condition
\begin{equation}
\dpart{f}{m} = 0 \Longrightarrow m = \tanh(\beta m), 
\end{equation}
which corresponds to the usual mean field equation for the magnetization. We also know that in the thermodynamic limit the system will thermalize inside one of the two states and remain there forever, so the presence of states induces an hard \emph{ergodicity breaking}\cite{pedestrians}.\\
Thus, we can see that in this very simple case everything is known: we know what the number of states is (two), we have an order parameter that allows us to tell apart one from the other, and most importantly, it is very easy to study ergodicity breaking: all that we have to do is apply an external magnetic field $h$, which couples to the magnetization $m$. Using the free energy, we then compute the magnetization as a function of the temperature $T$ and the magnetic field $h$. If we then send $h$ to zero, and obtain a nonzero magnetization, this will unambiguously mean that the two states have appeared and that ergodicity breaking has occurred; if this is not the case, then it will mean that the system is still in the ergodic phase.\\
This strategy is also viable for numerical simulations: we can introduce an external field to project the system onto one of the two states, lower the temperature below $T_c$, remove the field, and then compute the equilibrium order parameter from the simulation itself. If it happens to be nonzero, we will immediately know that ergodicity has been broken.

Now, let us move to a more complicated situation. We suppose to have a system wherein the magnetization is not homogeneous in space, but instead is lattice-site (or space) dependent
$$
m_i = \thav{\sigma_i},
$$
moreso, we complicate the problem even more by assuming that this spatial dependence is not ``regular'' in space, that is, it cannot be described by a ``simple'' function of the position\footnote{As it is, for example, in the anti-ferromagnetic Curie-Weiss model.}. We thus have a \emph{disordered} magnetization.\\
Now, the free energy will be a function of all those magnetizations $f(\{m_i\}_{i=1}^N)$\footnote{In the framework of spin models, this is called the Thouless-Anderson-Palmer(TAP) free energy \cite{TAP77}.}, and the stationarity condition will thus be given by $N$ different equations,
\begin{equation}
\dpart{f}{m_i}  = 0\qquad(i=1,\dots,N).
\label{disordequations}
\end{equation}
Models which exhibit this phenomenology can be constructed in a variety of ways. In spin models, one usually induces disorder artificially by assuming that the couplings between spins $J_{ij}$ are quenched random variables with a certain probability distribution \cite{pedestrians}, while in the case of structural glasses the disorder is self-induced by the system itself; however, this does not make any relevant difference; all that we are going to say about spin models can be carried over to structural glasses without difficulty, as we are going to see.

It is now easy to notice that this complication has completely disrupted the whole nice theoretical framework that we could rely on for the Curie-Weiss model: unless we are able to solve \emph{all} the $N$ equations for the magnetizations (with $N$ going to infinity, nonetheless), we cannot use the magnetizations to label states anymore, since, without a solution, we don't know which ``vector'' of magnetizations $\mathbf{m}_\alpha \equiv (\{m_i\}_{i=1}^N)_\alpha$ identifies each state. Besides, it is obvious that in this case we will have a large number of solutions instead of the just two that we had before: if all magnetizations are forced to be equal, the number of solutions is small and it can be determined easily just by looking at the symmetries of the Hamiltonian. In the disordered case, we have no symmetries and all magnetizations can be different, so it is quite obvious that the number of solutions (that is, of states) is going to be very large, as it is obvious that the number of states cannot 
be computed from 
the equations \refeq{disordequations}\footnote{This is not entirely true as it is indeed possible to compute analytically the number of states from the TAP free energy, at least for the p-spin spherical model (see \cite{cavagnagiardina98,pedestrians}). But this strategy is quite convoluted and not viable for structural glasses.}. Indeed, the situation is even worse since not only there are many states, but there are even many \emph{for each value of the state free energy}, that is, for each value of the free energy $f$ at the minimum linked to the state. So the states are not only many, but they can also be \emph{degenerate}.\\
Another important setback is the fact that now we have lost the magnetic field $h$ as a crucial tool for selecting states. Of course, for each state $\mathbf{m}_\alpha$ there will be a disordered field $\mathbf{h}_\alpha$ that projects the system onto it. But since we don't know what $\mathbf{m}_\alpha$ is, we don't know $\mathbf{h}_\alpha$ either. Thus, we have lost the order parameter as a label for the states, we are unable to know how many states we have, and our strategy for the study of ergodicity breaking is now unusable.

\subsection{Real replicas}

This is where the replica method comes in. Suppose that we have a generic system (it can be a spin model, or a liquid, or whatever) which has the disordered properties that we just enumerated: lack of a simple order parameter and presence of many equivalent states\footnote{Up to now we have made no distinction between \emph{stable} and \emph{metastable} states; indeed, in a mean-field scenario, there is not that much of a difference since metastable states have infinite lifetime and are able to trap the dynamics exactly as stable states do. However, out of the mean-field the distinction is actually very important and must not be forgotten.}. To be general, let us assume that the system can be described by a coarse-grained field theory in the order parameter $\phi(\bx)$, with a generic Hamiltonian $H[\phi]$, and an external field $h(\bx)$ coupled with the order parameter via a small constant $\e$. The free energy at inverse temperature $\beta$ reads then
\begin{equation}
F_{\phi}[h(\bx);\e,\beta] = -\frac{1}{\beta}\log\int \delta\phi(\bx)\ \exp\left(-\beta H[\phi] -\frac{\epsilon}{2}\int d\bx\ [h(\bx)-\phi(\bx)]^2\right).
\label{monassonF}
\end{equation}
Let us suppose that we know a configuration of the field $h(\bx)$ such that a certain state (or equivalently a certain minimum of the free-energy landscape) is selected. When the coupling is nonzero, the field $h(\bx)$ will break the symmetry of $H[\phi]$ and force the field $\phi(\bx)$ to lie along its direction to minimize the free energy. If the model shows a phase transition, there will be a certain temperature $T_c$ below which the order parameter $\phi(\bx)$ will remain frozen in this certain configuration even at zero coupling ($\e \to 0$) because the symmetry breaking is spontaneous below $T_c$. Indeed, we can see that the only difference of this case with respect to the Curie-Weiss model lies in the fact that we do not know which is the right configuration for $h(\bx)$.\\
However, we do know a thing about it: we know that such a configuration would minimize the free energy \refeq{monassonF} for $\e\to0$. Following Monasson \cite{monasson}, we now assume that the field $h(\bx)$ is not ``external'' (in the sense that it is a fixed variable of the statistical ensemble we are considering, like the temperature), but rather a variable, thermalized at a certain inverse temperature $\beta'$, whose Hamiltonian is just the $F_{\phi}[h(\bx),\e,\beta]$. The free energy of the field $h(\bx)$ would then read
\begin{equation}
F_h(\beta') = \lim_{\e\to0} -\frac{1}{\beta'}\log \int \delta h(\bx)\ \exp\left(-\beta'F_{\phi}[h(\bx);\e,\beta]\right).
\end{equation}
So, since we do not know the right $h(\bx)$, we simply go and ask the system itself which one it is.\\
This free energy is impossible to compute in the general case. However, if we assume that
$$
\beta' = m\beta,
$$
where $m$ is an integer number, we get by definition of $F_\phi$
\begin{equation}
\begin{split}
F_h(\beta,m) = &\ \lim_{\e\to0} -\frac{1}{\beta m}\log \int \delta h(\bx)\left[\int \de\phi(\bx)\ \exp\left(-\beta H[\phi] -\frac{\epsilon}{2}\int d\bx\ [h(\bx)-\phi(\bx)]^2\right)\right]^m\\
=&\ \lim_{\e\to0} -\frac{1}{\beta m}\log \int \delta h(\bx)\delta \phi^1(\bx)\dots \delta \phi^m(\bx)\exp\left(-\beta \sum_{a=1}^m H[\phi^a] -\sum_{a=1}^m\frac{\epsilon}{2}\int d\bx\ [h(\bx)-\phi^a(\bx)]^2\right).
\end{split}
\end{equation}
We can now evaluate easily the free energy $F_h$, as we only have to perform a Gaussian field integration with a linear term. We finally get (modulo an infinite constant)
\begin{equation}
F_h(\beta,m) =\ \lim_{\e\to0} -\frac{1}{\beta m} \log\int \prod_{a=1}^m \de\phi^a(\bx)\exp\left(-\beta\sum_{a=1}^m H[\phi^a] -\sum_{a<b}^m\frac{\epsilon}{2m}\int d\bx\ [\phi^a(\bx)-\phi^b(\bx)]^2\right).
\label{replicatedF}
\end{equation}
The meaning of equation \refeq{replicatedF} is clear: we have to study the statics of a system made of $m$ weakly coupled \emph{replicas} of the original one.\\
Let us focus on the coupling term. We see that is is small when the fields corresponding to different replicas are similar, and it is big when they are decorrelated. So we have that the weak coupling $\e$ can be seen as a constant external field conjugated with a parameter which measures the ``distance'' between different replicas.\\
Generally, the definition of this distance will depend on the particular problem in study; apart from the field theoretic case, we can for example consider spin systems, in which case it is called \emph{overlap}, denoted as $q$, and defined as 
\begin{equation}
q \equiv \frac{1}{N}\sum_{i=1}^N \s_i^a \s_i^b,
\end{equation}
where $a$ and $b$ are indexes that label two different replicas; the overlap measures the average degree of correlation between the configurations of the two replicas, so it is actually a \emph{codistance}: the higher, the nearer. For glass forming liquids, wherein the degrees of freedom are positions and momenta of the particles, the distance is the \emph{cage radius} $A$, which is a real distance, in real space, between different replicas of the same original particle. We will return to this definition in the following.

To better illustrate how the study of the replicated free-energy \refeq{replicatedF} can give us information about ergodicity breaking, let us follow again \cite{monasson} in considering a system made up of $n$ groups of $m$ replicas, each of them coupled according to the \refeq{replicatedF}:
\begin{equation}
F_h(n,m,\beta,\e)=\ \lim_{\e\to0} -\frac{1}{\beta m} \log\int \prod_{a=1}^{mn} \de\phi^a(\bx)\exp\left(-\beta\sum_{a=1}^{mn} H[\phi^a] -\sum_{c=1}^{n}\sum_{a<b}^{a,b\in m_c}\frac{\epsilon}{2m}\int d\bx\ [\phi^a(\bx)-\phi^b(\bx)]^2\right).
\end{equation}
Notice again the ``distance'' parameters $q_{ab} \equiv \int d\bx\ [\phi^a(\bx)-\phi^b(\bx)]^2$. If $\e = 0$, the coupling vanishes and the Hamiltonian is perfectly symmetric under permutations of any couple of replicas. However, the presence of the coupling explicitly breaks this symmetry, as only replicas in the same group are coupled. This breaking of the permutation symmetry is referred to as \emph{one-step replica symmetry breaking} (1RSB).\\
But what if the breaking of replica symmetry were spontaneous? Let us consider the Gibbs free energy of the replicated system, that is, the thermodynamic potential at fixed $q_{ab}$, as opposed to fixed $\e$. To keep into account the explicit breaking of replica symmetry, we assume it to have the form
\begin{equation}
G(n,m,d_0,d_1),
\end{equation}
where $q_{ab} = d_1$ if the replicas $a$ and $b$ in the same group $m_c$, and $q_{ab} = d_0$ if they belong to different groups. This is the minimal form that allows us to keep RSB into account. The free energy $F_h$ at zero coupling will then be the Legendre transform of the $G$ with respect to $d_0$ and $d_1$ for $\e=0$, which means
\begin{equation}
F_h(n,m,\beta,\e=0) = \textrm{Ext}_{d_0,d_1}G(n,m,d_0,d_1).
\label{gibbs}
\end{equation}
So the free energy at zero coupling is computed by considering the extremum points of the Gibbs free energy with respect to the two order parameters, exactly as the free energy of the Curie-Weiss model at zero field is given by the $f(m)$ calculated on its stationary points.

Let us now suppose that we are in the simplest possible situation: perfect ergodicity, one state only. 
\begin{figure}[t!]
\begin{center}
\includegraphics[width=0.5\textwidth]{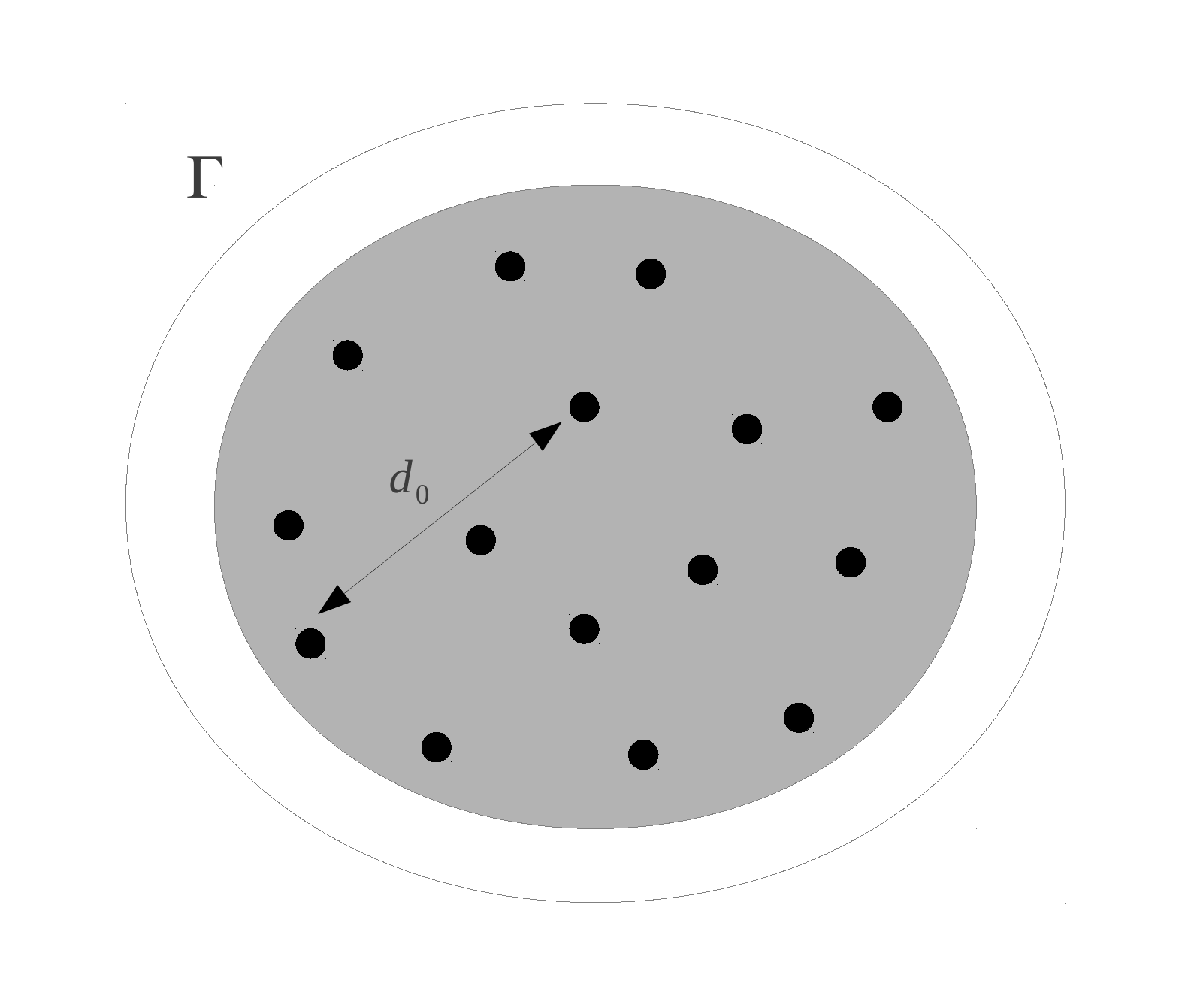}
\caption{Phase space in presence of only one ergodic state. Each replica corresponds to a black dot in the phase space $\Gamma$ of the original (nonreplicated) system. The grey blob corresponds to the only state.}
\label{RS}
\end{center}
\end{figure}
This situation is sketched in figure \ref{RS}. In this case, the replicas in the coupled groups, without the coupling $\e$ to keep them together, just scatter away in the phase space of the system. The distance between any two of the replicas is always the same and does not depend on the particular pair that we selected, so we have $d_0=d_1$: all replicas are perfectly equivalent and we can permute them as we please. We are then in a \emph{replica symmetric} (RS) scenario.

\begin{figure}[t!]
\begin{center}
\includegraphics[width=0.5\textwidth]{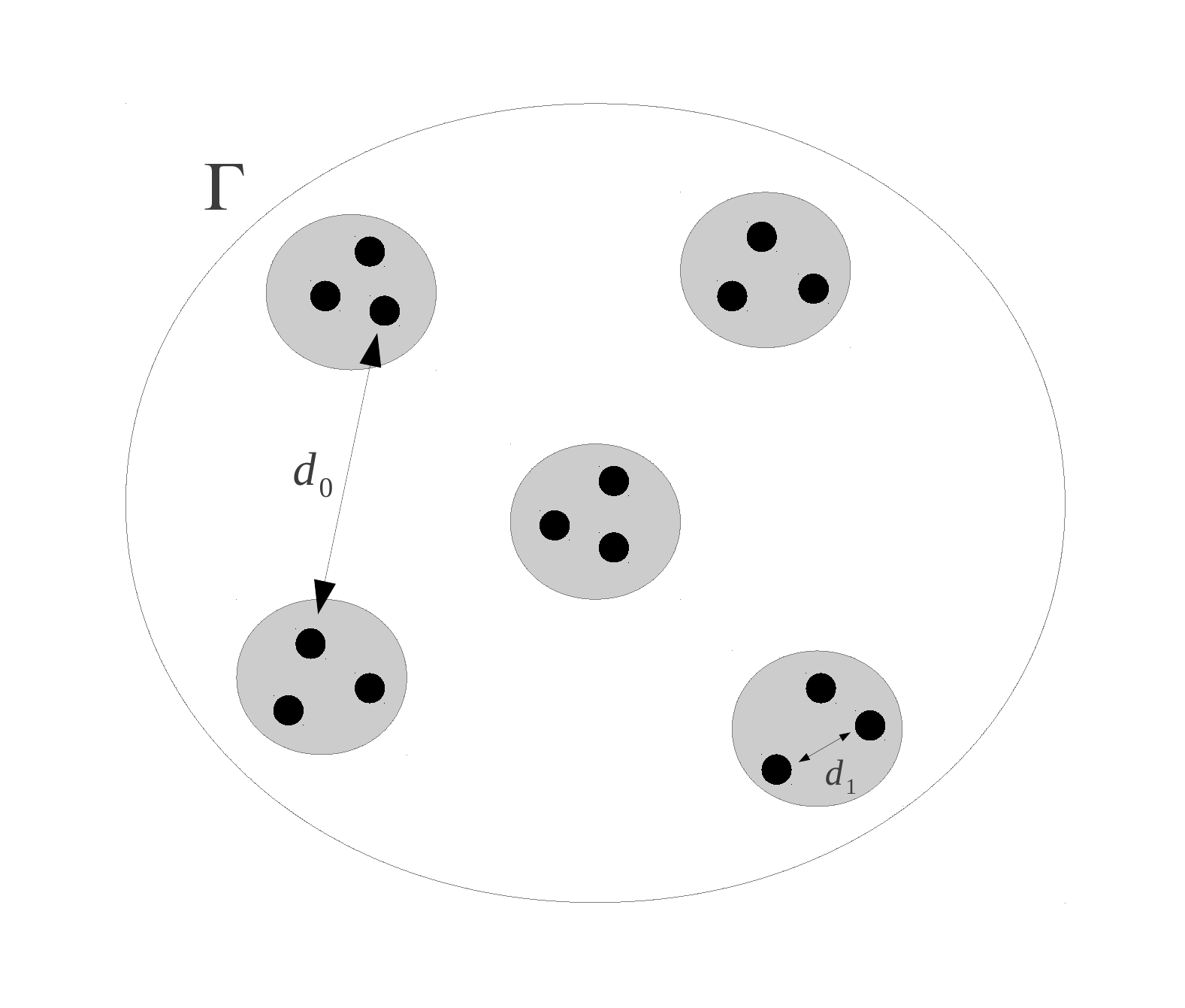}
\caption{Phase space in presence of many states. The replicas group inside states in such a way that $m$ of them are in each state.}
\label{1RSB}
\end{center}
\end{figure}

Now, we change the situation. We assume that ergodicity has been broken and different states are present, each corresponding to a blob of microscopic configurations\footnote{For a good discussion about the nature of states as basins of configurations, and a possible operative definition of them, see \cite[Appendix A]{berthierjacquin11}.} in the phase space of the original system; this situation is sketched in figure \ref{1RSB}.
We can now see that the picture has changed: the replicas in the groups stay near each other even for $\e=0$, because the state in which they found themselves keeps them together\footnote{Remember that replicas cannot get out of the states because of ergodicity breaking.}; it is then clear that replicas are no longer equivalent even for zero coupling: the distance between replicas belonging to different states $d_0$ will be different from the distance between replicas belonging to the same state $d_1$, and it is not true anymore that we can operate a permutation of the replicas as we please. The replica symmetry (that is, symmetry under permutation of the replicas) has been \emph{spontaneously} broken.

It is important to understand that this replica symmetry breaking plays exactly the same role as the breaking of the symmetry under spin reversal in the Curie-Weiss model. In both cases, the Hamiltonian is invariant under a certain symmetry, but we find out that the equilibrium state of the system is not. Indeed, many equilibrium states form, all linked by the symmetry and all equivalent, in the sense that despite being physically different, they all have the same free energy (the free energy for the replicated system is invariant under permutation of the replicas, in \emph{any case}). In both cases, the symmetry breaking signals us the birth of states.\\
This symmetry breaking can be iterated: for example, one can have states inside clusters of states, corresponding to a 2-RSB situation; then I can iterate again, getting states inside clusters of states inside metaclusters, and so on, getting a $k$-RSB structure. The process can go on indefinitely, and in fact, for the SK model and mean-field hard sphere glasses, it does! \cite{parisiSGandbeyond,fullRSB,corrado2}.\\  
Thus, in order to investigate ergodicity breaking, all we have to do is compute the values of $d_0$ and $d_1$ from the free energy of the replicated system (usually from an optimum condition) and check if there are any non-trivial (that is, different from $d_0$) solutions for $d_1$: when this happens, it will mean that states have formed and ergodicity breaking has occurred: this is the prescription for the study of the dynamical transition in the 1RSB replica method.

We must stress the fact that the structure of states cannot be determined \emph{a priori} and so one must take a guess (an \emph{ansatz}) as to which it might be. For the p-spin spherical model (PSM), for example, it is possible to check \emph{a posteriori} that the 1RSB solution is exact\cite{pedestrians}, but this is not the case both for the p-spin with discrete spins (at least for low enough temperature \cite{gardner85,grosskanter85}) and the SK model \cite{parisiSK} (that is, the 2-spin with discrete spins).\\
In the case of structural glasses, RFOT surmises that the structure of states is indeed 1RSB, basing this assumption on the analogy of their dynamics to the one of the PSM\cite{kirkpatrickthirumalai}, although recently this view has been proven wrong, at least for hard spheres at high enough densities\footnote{Indeed, the results of \cite{zamponiurbani,fullRSB} show that structural glasses are actually in the same universality class as the Ising p-spin, and not the PSM.}\cite{zamponiurbani,fullRSB}. However, the 1RSB ansatz is all that is needed for the study of low density regime and the dynamical transition.

\subsection{Complexity and internal entropy}

The replica method provides us with a way to investigate ergodicity breaking (i.e. the dynamical transition, for 1RSB systems), but we still need a way to compute the number of states; luckily, the replica method allows us to do that.
Let us go back to the system of $m$ coupled replicas, forced to be in the same state. We now specify out treatment to the case of a system of hard spheres, where temperature is irrelevant (in only enters the ideal-gas part of thermodynamic functions) and the entropy, instead of the free energy, is normally used as a thermodynamic potential \footnote{This happens because an hard-sphere system, once the kinetic (ideal gas) part of the Hamiltonian has been discarded, does not have any energy levels (only forbidden configurations due to the hard-core constraint) and so the canonical ensemble is equivalent to the micro-canonical one, making entropy the suitable thermodynamic potential.}. The control parameter will then be the \emph{packing fraction} $\varphi$, i.e., the fraction of volume occupied by the spheres: $\varphi \equiv \frac{Nv_d}{V} = \rho v_d$.\\
The partition function is:
$$
Z_m = (Z)^m = \left(\sum_\al e^{Ns_\alpha}\right)^m = \sum_{\al_1,\al_2,...\al_m} e^{N(s_{\al_1} + s_{\al_2} + \dots + s_{\al_m} )}, 
$$
where $s_\alpha$ is the intensive entropy of state $\alpha$, and for each replica, we have written the partition function by exploiting its decomposition into states. Now, since all replicas are in the same state because of the coupling, we have that $s_{\al_1} = s_{\al_2} = \dots = s_{\al_m}$, so 
$$
Z_m = \sum_\al e^{Nms_\al}.
$$
We can then rewrite the sum as an integral using delta functions:
$$
Z_m = \sum_\al e^{Nms_\al} = \int ds\ \sum_\al\delta(s-s_\al)e^{Nms},
$$
and then we define the \emph{complexity} $\Sigma(s,\varphi)$ in the following way,
\begin{equation}
\Sigma(\varphi,s) = \frac{1}{N}\log\left(\sum_\al \delta(s-s_\al)\right),
\end{equation}
i.e, it is the number of states which have entropy $s$, divided by the number of particles $N$. This way, the integral takes the form
$$
Z_m = \int ds\ e^{N(\Sigma(\varphi,s) + ms)}. 
$$
Now, since we are interested in the thermodynamic limit, we can evaluate the integral using the saddle-point method:
\begin{equation}
Z_m = \int ds\ e^{N(\Sigma(\varphi,s) + ms)} \simeq e^{N(\Sigma(\varphi,s^*) + ms^*)},
\end{equation}
and thus for the replicated entropy we have 
$$
\Sc(m,\varphi) = \frac{1}{N}\log Z_m = \Sigma(\varphi,s^*(m,\varphi)) + ms^*(\varphi,m)
$$
where $s^*(m,\varphi)$ is given by the stationarity condition
\begin{equation}
\left.\dtot{\Sc}{s}\right|_{s = s^*} = 0. 
\label{sstationary}
\end{equation}
As we can see, we now have an expression for the entropy of the replicated system in terms of the complexity and the internal entropy of states. It is now straightforward to check that
\begin{eqnn}
s^*(m,\varphi) &=& \dpart{\Sc}{m}, \label{vibrentropy}\\
\Sigma(m,\varphi) &=& \Sigma(s^*(m,\varphi),\varphi) = -m^2\dpart{(m^{-1}\Sc)}{m}; \label{complexity}
\end{eqnn}
indeed, since $s^*$ yields the stationarity condition \refeq{sstationary}, we don't need to account for the derivative $\dpart{\Sc}{s}\dtot{s}{m}$ because it is always zero. Once $s^*(m,\varphi)$ and $\Sigma(m,\varphi)$ are known, we can then reconstruct easily $\Sigma(s,\varphi)$ from their parametric plots as functions of $m$.\\
So, once we know the entropy of the replicated system, we can compute the complexity ad internal entropy of the states just by using the recipes above. The replica method has managed to solve (almost) all of the problems that the disorder had created. This set of tools can be easily translated to the case where the free energy is used as a thermodynamic potential \cite{zamponi}.

Now, our problem has shifted to computing the properties (that is, the thermodynamic potential) of a replicated glass-forming liquid. Then we must perform the analytic continuation of the resulting expression to real $m$, and compute the derivatives \refeq{complexity}. Before proceeding with this task, it is appropriate to review briefly the fundamentals of liquid theory.
\begin{figure}[h!]
\begin{center}
\includegraphics[width=0.5\textwidth]{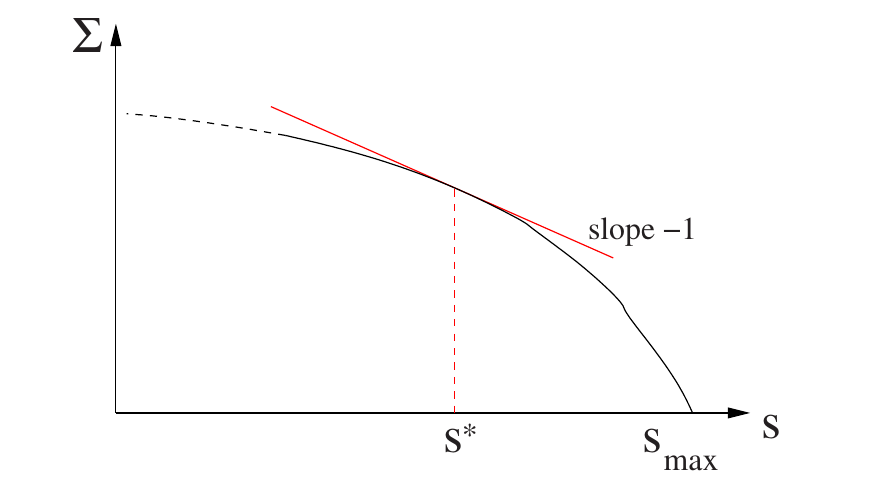}
\caption{Schematic representation of the complexity $\Sigma$ as a function of $s$. Reprinted from \cite{parisizamponi}.}
\end{center}
\end{figure}

\section{A glimpse of liquid theory}

For our purposes, a liquid\footnote{In this section we follow chapters 2 and 3 of \cite{simpleliquids}.} will be a system of particles with an Hamiltonian
$$
H = \sum_{i=1}^N \frac{\bp^2_i}{2m} + \sum_{i<j}V(\bx_i-\bx_j) = K(\bp^N) + V(\bx^N)
$$
where $V(\bx)$ is a suitable interaction potential; there are several possible choices for this potential, depending on the phenomenon of interest.\\ 
For what concerns glassy behavior, the hard-sphere (HS) potential is the simplest choice. It is defined in the following way,
$$
V(|\bx|) = \left\{\begin{array}{lcl} = 0 & & |\bx|>D \\ =\infty & & |\bx|<D\end{array}\right.
$$
where $D$ is the diameter of the particles. As we anticipated, with this choice the liquid has no potential energy levels: the potential energy is always zero for all configurations without overlaps between particles, while the configurations with at least two overlapping spheres have infinite energy and thus have zero probability, according to the Boltzmann-Gibbs distribution. 

The canonical distribution for the HS liquid is defined as usual,
$$
P(\bx^N,\bp^N) = \frac{1}{Z}e^{-\beta H(\bx^N,\bp^N)},
$$
where $Z$ is the partition function
$$
Z = \int\frac{d^N\bx d^N\bp}{N!} \ e^{-\beta H(\bx^N,\bp^N)}.
$$
The $\frac{1}{N!}$ factor is needed to ensure the correct counting of microscopic states: since particles are indistinguishable, all configurations that are linked by an arbitrary permutation of particle labels must be counted only once; without this caution, we would get paradoxical results in the form of a nonextensive entropy\footnote{In the following, we will introduce a model wherein particles are not indistinguishable, and we will see how this causes some problems.}.\\ 
The partition function can be written as
\begin{equation}
Z = \frac{Z_N}{N!\Lambda^{3N}},
\label{Z}
\end{equation}
where $\Lambda$ is the De Broglie thermal wavelength and $Z_N$ is the \emph{configuration integral}
\begin{equation}
Z_N = \int d^N\bx\ e^{-\beta V(\bx^N)}.
\end{equation}

\subsection{Particle densities and distribution functions}
Fundamental objects in liquid theory are the \emph{particle densities} and the corresponding \emph{distribution functions}. The \emph{equilibrium $n$-particle density} is defined as 
\begin{equation}
\begin{split}
\rho(\bx^n) =& \frac{N!}{(N-n)!}\frac{1}{Z}\int\frac{d^{(N-n)}\bx d^N\bp}{N!}\ e^{-\beta H(\bx^N,\bp^N)},\\
=& \frac{N!}{(N-n)!}\frac{1}{Z_N}\int d^{(N-n)}\bx\ e^{-\beta V(|\bx|)}.
\end{split}
\label{partdens}
\end{equation}
This quantity gives the average number of $n$-tuples of particles which have spatial coordinates in the $n$-dimensional volume element $d^n\bx$, irrespective of the coordinates of the other particles and of all momenta.\\
Indeed, it is immediate to see that if we integrate over the volume, we get the total number of $n$-tuples in the system,
$$
\int d^n\bx \rho(\bx^n) = \frac{N!}{(N-n)!},
$$
and of course for $n=1$ we get
$$
\int d\bx \rho(\bx) = N.
$$
The function $\rho(\bx)$ is the local density, or \emph{density profile} of the liquid; because of translational invariance, it is always constant in the liquid phase, and because of the normalization condition, we have that
\begin{equation}
\rho(\bx) = \rho = \frac{N}{V},
\end{equation}
where $V$ is the volume. So the density profile for a liquid is equal to the number density, unless the liquid lies in some external, non-homogeneous field.

We now introduce an useful representation for particle densities. Let us compute the ensemble average of the function
$$
\de(\bx-\bx_1);
$$
we have
\begin{equation}
\begin{split}
\thav{\de(\bx-\bx_1)} =& \frac{1}{Z} \int\frac{d^N\bx d^N\bp}{N!} \ \de(\bx-\bx_1) e^{-\beta H(\bx^N,\bp^N)}\\
=& \frac{1}{Z_N}\int d^N\bx\ \de(\bx-\bx_1) e^{-\beta V(\bx_1,\bx_2,\dots,\bx_N)}\\
=& \frac{1}{Z_N}\int d^{(N-1)}\bx\ e^{-\beta V(\bx,\bx_2,\dots,\bx_N)}.\\
\end{split}
\end{equation}

If we used $\bx_2$ instead of $\bx_1$, or any other particle, the result would not be different since all particles are equivalent. So, if we sum over all particles, we get
$$
\thav{\sum_{i=1}^N\de(\bx-\bx_i)} = N\frac{1}{Z_N}\int d^{(N-1)}\bx\ e^{-\beta V(\bx,\bx_2,\dots,\bx_N)},
$$
which is exactly $\rho(\bx)$. This way, we can see that the density profile indeed corresponds to the ensemble average of the microscopic particle density.\\ 
It is easy to generalize this representation to the higher-order particle densities. For example, the pair density $\rho(\bx,\by)$ can be written as
$$
\thav{\sum_{i\neq j}^N\de(\bx-\bx_i)\de(\by-\bx_j)} = \rho(\bx,\by).
$$

The \emph{$n$-particle distribution function} is defined in terms of the corresponding $n$-particle density in the following way:
\begin{equation}
g(\bx_1,\dots,\bx_n) \equiv \frac{\rho(\bx_1,\dots,\bx_n)}{\rho(\bx_1)\rho(\bx_2)\dots\rho(\bx_n)}.
\label{partdistrib}
\end{equation}
The full hierarchy of $n$-particle distribution functions encodes the structure of the liquid and particle correlations, and knowing it is equivalent to completely solving the statics of the liquid. However, when the potential is pairwise additive (as it is often the case), one can see that all of the static quantities of interest (internal energy, pressure, equation of state, etc...) can be computed from knowledge of the pair distribution function only:
$$
g(\bx,\by) = \frac{\rho(\bx,\by)}{\rho^2}.
$$
If the fluid is homogeneous and isotropic, then the $g(\bx-\by)$ is a function of the particle separation only, $g(r = |\bx-\by|)$, and is called the \emph{radial distribution function}. Pretty much all approximations in liquid theory, starting from the classic ones, Hyper-Netted Chain (HNC) and Percus-Yevick (PY), focus on the computation of the radial distribution $g(r)$ \cite{simpleliquids}.

\subsection{Internal energy and pressure}
As an example, we now sketch rapidly how it is possible to compute internal energy and pressure of the liquid from the $g(r)$. We start from the basic definition for the internal energy
$$
U = \thav{H} = \frac{1}{Z}\int \frac{d^{N}\bx d^{N}\bp}{N!} (K(\bp^N) + V(\bx^N))e^{-\beta(K(\bp^N) + V(\bx^N))};
$$
the average of the kinetic term is trivial and gives the ideal gas internal energy
$$
U_{id} = \frac{3}{2}Nk_{B}T,
$$
while the average of the potential term reduces to\footnote{The pedix ``ex'' stays for ``excess'', which is standard terminology in liquid theory.}
$$
U_{ex} = \frac{1}{Z_{N}}\int d^{N}\bx\left[\sum_{i<j}V(r_{ij})\right]e^{-\beta V(\bx^N)},
$$
with the definition
$$
r_{ij} \equiv |\bx_{i}-\bx_{j}|.
$$
Because of the symmetry of the problem under permutation of particle labels, each term of the sum has the same value, so we can write
$$
U_{ex} = \frac{N(N-1)}{2}\int d\bx_{1} d\bx_{2}V(r_{12}) \left(\frac{1}{Z_{N}}\int d\bx_{3}\dots d\bx_{N}\ e^{-\beta V(\bx^N)}\right),
$$
and then, using definitions \refeq{partdens} and \refeq{partdistrib}, we get
$$
U_{ex} = \frac{1}{2\rho^{2}}\int d\bx_{1} d\bx_{2}\ g(\bx_{1},\bx_{2})V(r_{12});
$$
then, for homogeneous, isotropic fluids, where the $g$ depends only on the separation $\bx_{1}-\bx_{2}$, we can change coordinates defining $\bx \equiv \bx_{1}-\bx_{2}$ and integrate over $\bx_{2}$ (which gives a factor $V$ at the numerator), and we get
\begin{equation}
\frac{U_{ex}}{N} = u_{ex} = \frac{\rho}{2}\int d\bx\ g(r)V(r) = 2\pi\rho\int_{0}^{\infty} dr\ r^{2}g(r)V(r),
\end{equation}
which shows us that knowledge of the radial distribution function alone is enough to compute the excess internal energy per particle of the liquid.\\
To compute the pressure, one starts from Clausius' \emph{virial function}, defined as
\begin{equation}
\mathcal{V}(\bx^{N}) \equiv \sum_{i=1}^{N} \bx_{i}\cdot \mathbf{F}_{i}(\bx^{N}).
\end{equation}
The force contained in the virial function can be separated in two parts: an internal part exerted by other particles, and thus linked to the potential energy, and external part which is exerted by the walls of the container, and is thus linked to the pressure. The force $\mathbf{dF}$ exerted by a surface element $dS$ of the container wall is by definition equal to $-P\ \mathbf{n}dS$ (where $\mathbf{n}$ is the unit vector orthogonal to the surface element), so the average of the external part of the virial can be written as
$$
\thav{\mathcal{V}_{ext}} = -P\int\bx\cdot\mathbf{n}\ dS = -P\int \nabla\cdot\bx\ d\bx = -3PV,
$$
where we have used the divergence theorem. Then, using the virial theorem
$$
\thav{\mathcal{V}}= \thav{\mathcal{V}_{int}}+\thav{\mathcal{V}_{ext}} = -2\thav{K} = -3Nk_{B}T,
$$
one can get the \emph{virial equation}
\begin{equation}
\frac{\beta P}{\rho} = 1-\frac{\beta}{3N}\left<\sum_{i=1}^{N}\sum_{j\neq i}^{N}\bx_{i}\cdot\nabla_{i} V(r_{ij})\right>.
\end{equation}
Now, the sum between brackets can be rewritten as
$$
\sum_{i=1}^{N}\sum_{j\neq i}^{N}\bx_{i}\cdot\nabla_{i} V(r_{ij}) = \sum_{i=1}^{N}\sum_{j<i}^{N}\bx_{i}\cdot\nabla_{i} V(r_{ij}) + \sum_{j=1}^{N}\sum_{i<j}^{N}\bx_{j}\cdot\nabla_{j} V(r_{ij}),
$$
then we can use Newton's third law and get
$$
\sum_{i=1}^{N}\sum_{j\neq i}^{N}\bx_{i}\cdot\nabla_{i} V(r_{ij}) = \sum_{j<i}r_{ij}V'(r_{ij}),
$$
where with $V'(r)$ we denote the derivative of the interaction potential $V(r)$ with respect to $r$.\\
Now, we can compute the average of this function following the same exact steps as we did for the excess free energy, and in the end we get
\begin{equation}
\frac{\beta P}{\rho} = 1-\frac{2\pi\beta\rho}{3}\int_{0}^{\infty}dr\ r^{3}g(r)V'(r).
\label{virial}
\end{equation}
This allows us, once the $g(r)$ is known, to compute the pressure of the liquid, and from it, the equation of state. These two examples illustrate well why the $g(r)$ is, in many situations, the central object in liquid theory, more than the partition function or the thermodynamic potential.\\
In the case of hard spheres, equation \refeq{virial} cannot be applied as it is since the interaction potential is actually discontinuous. However, we can circumvent this difficulty by defining another function, the \emph{cavity distribution function}
$$
y(r) \equiv e^{\beta V(r)}g(r).
$$
we can then plug this definition in the \refeq{virial}, getting
\begin{equation}
\begin{split}
\frac{\beta P}{\rho} &= 1-\frac{2\pi\beta\rho}{3}\int_{0}^{\infty}dr\ r^{3}V'(r)e(r)y(r).\\
&= 1+ \frac{2\pi\rho}{3}\int_{0}^{\infty}dr\ r^{3}e'(r)y(r)
\end{split}
\end{equation}
where we have defined
$$
e(r) \equiv e^{-\beta V(r)};
$$
for the hard sphere potential, as it is immediate to check, this function corresponds to a Heaviside theta,
$$
e(r) = \theta(r-D),
$$
and thus we have
$$
e'(r) = \delta(r-D)
$$
and so, in the end, we get
\begin{equation}
\frac{\beta P}{\rho}  = 1+ \frac{2\pi\rho}{3}\lim_{r\to D^+ } r^{3}y(r) = 1+ \frac{2\pi\rho}{3}\ D^{3}g(D).
\end{equation}
Thus, for HS the pressure is proportional to the value of the pair distribution function at contact. This makes so that the radial distribution function for an HS liquid is a central tool in the study of the \emph{jamming transition} \cite{parisizamponi,fullRSB}, whose main signature is indeed the divergence of the pressure.

\subsection{Diagrammatic methods}
We conclude this section with a very brief account of the commonly used methods for the computation of the static properties of a liquid.\\
Usually, the partition function \refeq{Z} cannot be computed exactly for all but the simplest Hamiltonians, thus making necessary the use of approximate methods. Apart from the classic approximations of liquid theory (HNC and PY) which are devised to compute the $g(r)$ as a solution of certain integral equations, another fruitful method relies on diagrammatic expansions\footnote{Indeed, the HNC approximation itself can be derived and (somehow) justified diagrammatically.} for the thermodynamic potential of interest ($\Omega$ in the case of the grancanonical ensemble, $F$ in the case of the canonical ensemble). Once the thermodynamic potentials are known, the distribution functions can be recovered as their functional derivatives with respect to either the local activity (for the grancanonical ensemble) or the local density profile (for the canonical one). 

\begin{figure}[htb!]
\begin{center}
\includegraphics[width=0.35\textwidth]{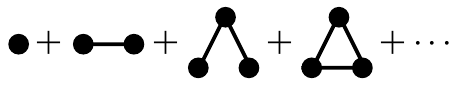}
\caption{Diagrams for the $\Omega[z^{*}(\bx)]$, up to the third order in the activity.}
\label{diagrams1}
\end{center}
\end{figure}

As an example, we provide the diagrammatic expansion for the Landau potential $\Omega$, as a functional of the local activity $z^{*}(\bx)$
$$
z^{*}(\bx)\equiv e^{\beta\mu-\beta\phi(\bx)},
$$
where to be general we assume that the liquid lie in some external field $\phi(\bx)$. The Landau potential can then be expressed diagrammatically as
\begin{equation}
-\beta\Omega[z^{*}(\bx)] = [\textrm{Sum of all connected diagrams with $z^{*}$-vertices and $f$-bonds}],
\label{Omega}
\end{equation}
where $f(\bx-\by)$ is the \emph{Mayer function},
$$
f(\bx-\by)\equiv e^{-\beta V(\bx-\by)} -1.
$$
As an example, in figure \ref{diagrams1} we show the first diagrams for the $\Omega$.
To compute one diagram, one must assign a label variable to each vertex and then evaluate the corresponding integral. The result has then to be divided by the \emph{symmetry factor} $S$ of the diagram, that is, the number of relabeling of the vertices that leave the connections unaltered. For example, the integral corresponding to the third diagram in figure \ref{diagrams1} is
$$
\int d\bx_{1}d\bx_{2}d\bx_{3}\ z^{*}(\bx_{1})f(\bx_{1}-\bx_{2})z^{*}(\bx_{2})f(\bx_{2}-\bx_{3})z^{*}(\bx_{3}),
$$
and the symmetry factor is equal to 2: One can switch vertex 1 with vertex 3, but cannot exchange, say, 2 with 3, because 2 has remain connected to 1.

\begin{figure}[htb!]
\begin{center}
\includegraphics[width=0.5\textwidth]{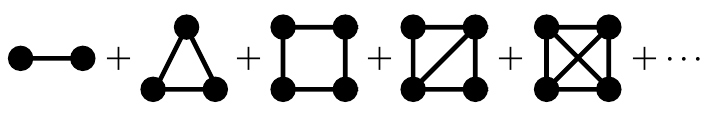}
\caption{Diagrams for the $F[\rho(\bx)]$, up to the fourth order in the density.}
\label{diagrams2}
\end{center}
\end{figure}

\begin{figure}[htb!]
\begin{center}
\includegraphics[width=0.45\textwidth]{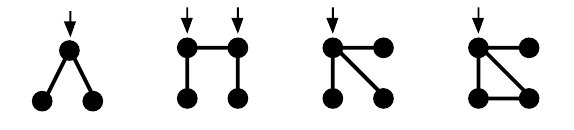}
\caption{Some diagrams with articulation vertices.}
\label{irreducible}
\end{center}
\end{figure}

From the diagrammatic expression of $\Omega[z^{*}(\bx)]$, one can obtain, with suitable diagrammatic operations (see \cite{simpleliquids} for details), an analogous expression for $F[\rho(\bx)]$, which reads:
\begin{equation}
\begin{split}
-\beta F[\rho(\bx)] = &\int d\bx\ \rho(\bx)[1-\log(\Lambda^{3}\rho(\bx))] + \\
&[\textrm{Sum of all irreducible diagrams with $\rho$-vertices and $f$-bonds,}]
\label{F}
\end{split}
\end{equation}
where for ``irreducible'' diagram we mean a diagram free of articulation vertices, i.e., vertices whose removal makes the diagram disconnected (figure \ref{irreducible}). The first diagrams are in figure \ref{diagrams2}. The passage from $\Omega$ to $F$ is equivalent to taking a functional Legendre transform of $\Omega$ with respect to the ``variable'' $z^{*}(\bx)$ conjugated to $\rho(\bx)$.

As we said before, the distribution functions (and the particle densities, by extension) can be expressed in terms of functional derivatives of the thermodynamic potentials. For example, for the $n-$particle density we have
$$
\rho(\bx_{1},\bx_{2,},\dots,\bx_{n}) = \frac{z^{*}(\bx_{1})\dots z^{*}(\bx_{N})}{\mathcal{Z}}\frac{\de^{n}\mathcal{Z}}{\de z^{*}(\bx_{1})\dots \de z^{*}(\bx_{N})},
$$
where 
$$
\mathcal{Z}[z^{*}(\bx)] = e^{-\beta\Omega[z^{*}(\bx)]}.
$$
Thus, one can obtain similar diagrammatic expansions for the particle densities by functional differentiation of the expression \refeq{Omega} for $\Omega$.

\section{How is a glass similar to the disordered ferromagnet?}

Let us take a step backwards. In the preceding section, we mentioned the fact that in the liquid phase the density profile is a constant in space, equal to the number density
$$
\rho(\bx) = \rho,
$$
exactly like the magnetization in the Curie-Weiss model we mentioned in the first section. In this sense, a liquid can be seen as a sort of ``paramagnet''.\\
But what does happen when the liquid becomes a glass?\\
What happens is that the density profile is not constant anymore, but instead, because of the amorphous structure of the glass, density fluctuations are now allowed and so $\rho(\bx)$ becomes a function of the position, exactly like the magnetization in the disordered ferromagnet. Indeed, the analogy can be pushed further as the density profile con be seen as a solution of a minimum condition on the free energy functional $F[\rho(\bx)]$\footnote{This statement is a fundamental theorem in density functional theory, the Hohenberg-Kohn-Mermin theorem, which states that the free energy is a functional of the 1-particle density profile \emph{only}, and that the equilibrium density profile is the one which minimizes the F \cite{simpleliquids}.}
\begin{equation}
\dfunc{F}{\rho(\bx)} = 0
\label{eqrho}
\end{equation}
exactly like the $m_{i}$ were. Besides this, the function $\rho(\bx)$, since it describes an amorphous solid, cannot have a somehow ``simple'', ordered structure like it would in the case of a crystalline solid, wherein the density field would be a periodic function of the position,
$$
\rho(\bx) = \rho(\bx+\mathbf{R}).
$$
So, we have that, from this point of view, the glass is analogous to the disordered ferromagnet, and the crystalline solid to the antiferromagnet. Once the proper conceptual links are made, we find ourselves in the exact same situation as before.\\
Thus, in order to study the glass state, we should construct the energy functional $F$, then find all possible non-constant and non-periodic solutions of the \refeq{eqrho} and count them in order to compute the total number of amorphous structures (i.e., of states\footnote{Every state in a real glass can be thought of as the set of all configurations covered by the system as the particles vibrate around equilibrium positions arranged in an amorphous structure. Thus, every state corresponds to such a structure.}). Even more than in the case of the disorder ferromagnet, it is evident that this strategy is basically a lost cause. Thus, to solve our problem, we turn, finally, to replicas.

\section{The replicated HS liquid}
We now tackle the problem enunciated at the end of section \ref{sec:1}, i.e. computing the static properties of an HS liquid made up of $m$ weakly coupled replicas. We start from the model Hamiltonian
\begin{equation}
H = \sum_{a=1}^{m}\sum_{i<j}V_{HS}(\bx_{i}^{a}-\bx_{j}^{a}) + \frac{\epsilon}{m}\sum_{i=1}^{N}\sum_{a<b}(\bx_{i}^{a}-\bx_{i}^{b})^{2},
\label{Hrepl}
\end{equation}
where we have discarded the kinetic part, as it gives the trivial, ideal-gas part of the thermodynamic functions. For convenience reasons, we have chosen the inter-replica potential to be harmonic, governed by a small coupling constant $\epsilon$.  We must stress the fact that, despite the fact that we started from replica formalism, which in the context of liquid theory can look somewhat exotic, the Hamiltonian \refeq{Hrepl} has nothing special: it is just the Hamiltonian for a molecular liquid, where each molecule is formed by particles belonging to different replicas, interacting via a perfectly harmonic potential; in the context of liquid theory, it is a pretty standard and straightforward problem.

We now want to compute the entropy of this molecular liquid. We start from the grancanonical partition function $\mathcal{Z}_{m}(\epsilon)$ for the model:
$$
\Z_{m}(\e) = \sum_{N=0}^{\infty}e^{\beta\mu}\int \frac{d^{N}\bx_{1}\dots d^{N}\bx_{m}}{N!} \prod_{i<j}\chi(\bx_{i}^{a}-\bx^{a}_{j})\prod_{i=1}^{N}\exp\left(-\frac{1}{m}\sum_{a<b}(\bx_{i}^{a}-\bx_{i}^{b})^{2}\right)
$$
where we have defined the function
$$
\chi(\bx-\by) \equiv \theta(|\bx-\by|-D).
$$
We can then define a sort of ``molecular position'' $\ox \equiv (\bx_{1},\bx_{2},\dots\bx_{m})$, which allows us to rewrite the partition function in very compact form
\begin{equation}
\Z_{m}(\epsilon) = \sum_{N=0}^{\infty} \int \frac{d^{N}\ox}{N!}\prod_{i=1}^{N}z^{*}(\ox_{i})\prod_{i<j}\oc(\ox_{i},\ox_{j}),
\label{finalZ}
\end{equation}
with the definitions
\begin{eqnn}
z^{*}(\ox) &\equiv& z\exp\left(-\frac{\epsilon}{m}\sum_{a<b}(\bx^{a}-\bx^{b})^{2}\right),\\
\oc(\ox,\oy) &\equiv& \prod_{a=1}^{m}\chi(\bx^{a}-\by^{a}).
\end{eqnn}
We then define the thermodynamic potential for the replicated HS liquid,
$$
\Sc(m,\varphi,\epsilon) \equiv \frac{1}{\thav{N}} \log\Z_{m}(\epsilon),
$$
where $\varphi$ is the packing fraction again.

Now, let us compute this quantity ($d$ is the number of spatial dimensions):
$$
-\frac{1}{(m-1)d}\dtot{\Sc(m,\varphi,\e)}{\e}.
$$
We have that
\begin{equation}
\begin{split}
-\frac{1}{(m-1)d}\dtot{\Sc(m,\varphi,\e)}{\e} =& \frac{1}{m(m-1)d\thav{N}} \left<\sum_{i=1}^{N}\sum_{a<b}(\bx_{i}^{a}-\by_{i}^{b})^{2}\right>\\
=&\frac{1}{2d\thav{N}} \left<\sum_{i=1}^{N}(\bx_{i}^{a}-\by_{i}^{b})^{2}\right>.
\end{split}
\end{equation}
Thus, the derivative of $\Sc$ with respect to $\e$ is a quantity which measures the average distance between particles belonging to (any, as replicas are all equivalent and can be permuted at will) two different replicas: it is exactly the ``distance'' we were looking for! The quantity is called the \emph{cage radius} and is denoted as $A$.\\
We can again appreciate the remarkable analogy with the magnetic system: just as the magnetization is equal to the derivative of the free energy with respect to the magnetic field $h$, the cage radius is proportional the derivative of the potential $\Sc$ with respect to the coupling $\e$, which is the external ``field'' that constrains the replicas to be in the same state. Just as the appearance of a spontaneous magnetization at zero magnetic field was the signature of the ferromagnetic transition, now the appearance of a finite $A$ for zero coupling will signal the fact that states have appeared, as anticipated in section \ref{sec:1}.

At this point, it really looks like we are holding all the cards: computing the $\Sc$ is now a standard liquid theory problem. Since the partition function \refeq{finalZ} is written in the usual form for a liquid in the grancanonical ensemble (just with the molecular position in the place of the standard one), one could in principle compute it using, for example, the perturbative expansion \refeq{Omega}, having the caution to replace all the objects (activity, mayer function) with their molecular counterparts defined above. Once this has been done, we can study the behavior of $A$ at zero coupling, as a function of $\varphi$, and draw the phase diagram for the model. The complexity and free energy can then be computed easily from the $\Sc$ using the prescriptions \refeq{complexity} and \refeq{vibrentropy}.

However, in practice (and again, exactly as it is done for magnetic systems) it is more convenient to Legendre-transform $\Sc(m,\varphi,\e)$ with respect to $\e$, switching to a potential $\Sc(m,\varphi,A)$ which is an explicit function of $A$\footnote{One can also compute the entropy of the replicated liquid by generalizing in a suitable way the HNC and PY approximations. Historically, the replicated HNC approximation was the first method used to compute the properties of a replicated liquid, by M\'ezard and Parisi in 1996\cite{ParisiMezard}. However, one can see that for high enough $\varphi$ the RHNC performs very poorly.}
$$
\Sc(m,\varphi,\e) = \min_{\e}[\Sc(m,\varphi,\e)+d(m-1)A\e],
$$
in the same way as one passes from the Helmholtz free energy $F(\beta,h)$ to the Gibbs free energy $G(\beta,m)$ for a magnetic system. In order to do this, it is more convenient to switch to the canonical formalism from the very start (that is, use expansion \refeq{F}), rather than compute $\Sc$ perturbatively in the grancanonical formalism and then Legendre-transform to the canonical one. Once $\Sc(m,\varphi,A)$ is known, one can then compute $A$ from a minimum condition.\\
This however presents us with a problem: what is the ``molecular'' density profile $\rho(\ox)$, which we need to know in order to compute the diagrams in expansion \refeq{F}? In the canonical formalism, the field $\rho(\ox)$ is not given and is assumed to be known, in the exact same way in which the number of particles $N$ is an independent variable in the canonical ensemble, while it is not in the grancanonical one.\\ 
Thus, it becomes necessary to impose a certain form of the molecular density field, which must contain $A$ as a parameter of sorts. Let us look closely at the definition of $\rho(\ox)$:
$$
\rho(\ox) = \thav{\sum_{i=1}^{N}\delta(\ox-\ox_{i})} = \thav{\sum_{i=1}^{N}\prod_{a=1}^{m}\de(\bx^{a}-\bx_{i}^{a})}.
$$
We can thus notice that the molecular density field is really a sort of $m$-particle density, which is related to the probability of finding the $m$ replicas of particle $i$ (remember that only replicated particles with the same label $i$ interact via the harmonic potential) at positions $\bx_{a}, \bx_{b}, \dots, \bx_{m}$. Thus, it is the function which really describes the actual, physical shape of a molecule in the replicated liquid.

Knowing this, the simplest ansatz that we can make for the $\rho(\ox)$ is the following
\begin{equation}
\begin{split}
\rho(\ox) = &\frac{\rho m^{-d/2}}{2\pi A^{(m-1)d/2}}\exp\left(-\frac{1}{2mA}\sum_{a<b}(\bx_{a}-\bx_{b})^{2}\right),\\
= &\frac{\rho}{2\pi A^{md/2}}\int d\bX\ \exp\left(-\frac{1}{2A}\sum_{a=1}^{m}(\bx_{a}-\bX)^{2}\right),\\
=& \rho\int d\bX\ \prod_{a=1}^{m}\gamma_{A}(\bx_{a}-\bX),
\end{split}
\label{Gaussian}
\end{equation}
where $\gamma_{A}$ denotes a normalized Gaussian with variance $A$. As we can see, this ansatz corresponds to assuming that the displacements between particles in the molecule follow a Gaussian law with variance $A$. For this reason it is commonly called the \emph{Gaussian ansatz}.

Once we have this form for the molecular density profile, we just have to plug it in the diagrammatic expansion $\refeq{F}$ to compute the $\Sc(m,\varphi,A)$\footnote{The prescription is $\Sc[\rho(\ox)] = -\beta F[\rho(\ox)]$.}; then we can use the minimum condition \refeq{eqrho} and extremize the functional $F$ inside the Gaussian ansatz to get an equation for $A$
$$
\dtot{\Sc}{A} = 0,
$$ 
and search for any nontrivial solutions for this equation. It looks easy on paper.\\
Sadly, it is not. Computing the diagrams of the \refeq{F} with the Gaussian ansatz is indeed very complicated even at the second order, and even in that case it is not possible to have an analytic expression for every number of spatial dimensions, as we are going to see.\\
There are three possible ways out. The first is the \emph{small cage approximation} \cite{parisimezardjchem99,parisimezardprl99,parisizamponi}, in which the replicated liquid is replaced by an ordinary, atomic one with effective interaction potentials that can be computed, in powers of $\sqrt{A}$, from the diagrams that appear in the \refeq{F}. Therefore, for $A$ sufficiently small, one can hope to get acceptable results by considering a small number of interaction potentials (in fact, only the first one) \cite{parisizamponi}. It is a very cumbersome method, but for the moment is the only one that can be applied to the ``realistic'' model, i.e. the one defined by the Hamiltonian \refeq{Hrepl} in three dimensions.\\
The second is to consider a liquid embedded in a space with an \emph{infinite} number of spatial dimensions. In this case, it is possible to prove that the perturbative series \refeq{F} reduces to the first two terms only, i.e. the ideal gas term and the first interaction term (see \cite{frischpercus} for details). One can then extract the asymptotic results for high $d$. This method actually makes possible to compute things exactly, and it has been applied with remarkable success in the series of papers \cite{parisikurchan,zamponiurbani,zamponifull,fullRSB}. However, it has a big drawback in the fact that particle systems with an high number of spatial dimensions are hard to simulate numerically. Furthermore, an infinite dimensional system is unrealistic and it is hard to tell which results are still valid for 3-dimensional systems.\\
The third way, which is the one we are going to follow, is to switch from the usual HS model to a modified, \emph{mean-field} model, for which the perturbative series \refeq{F} reduces to the first two terms only without the need to go to infinite $d$. In the next section we will introduce and study briefly such a model.

\section{Explicit computations: the Mari-Kurchan (MK) model}
The Mari Kurchan model\cite{marikurchan} is defined by the Hamiltonian
\begin{equation}
H_{MK} = \sum_{i<j} V(\bx_{i}-\bx_{j}-\bA_{ij}),
\end{equation}
where $V$ is a suitable interaction potential (which in our case will be the HS one). The $\bA_{ij}$ are ``random shifts'', i.e. quenched, random $d$-dimensional vectors identically, independently and uniformly distributed in the $d$-dimensional cube:
$$
P(\bA) = \frac{1}{V}
$$
we also impose that $\bA_{ij} = \bA_{ji}$, for convenience reasons.

This model can be seen as ``mean field'' in multiple ways. First, we can notice that the model is devoid of any space structure: despite the fact that every particle interacts, given a certain realization of the $\bA$s, with a finite number of ``neighbors'' (so the model is not ``fully connected'' in the usual sense), those neighbors can be anywhere in the sample, since the shifts are uniformly distributed in the whole cube. From this point of view, the model is ``mean field'' because the physical space the model is embedded in plays no role on the interactions.\\
A less intuitive, but more profound line of reasoning stems from considering the probability of having three particles, say $i$, $j$ and $k$, interact with each other \emph{at the same time}, i.e., each of them interacts with \emph{both the other two} at the same time. For this to happen, we should have, for the HS potential,
\begin{equation}
\begin{split}
|\bx_{i}-\bx_{j} - \bA_{ij}|\simeq &\ D,\\
|\bx_{j}-\bx_{k} - \bA_{jk}|\simeq &\ D,\\
|\bx_{k}-\bx_{i} - \bA_{ki}|\simeq &\ D,
\end{split}
\end{equation}
which would imply
$$
|\bA_{ij} + \bA_{jk}+\bA_{ki}| \simeq D,
$$
which is very unlikely (and, in the thermodynamic limit, outright impossible), since the shift are $O(L)$\footnote{$L$ is the side of the cube.}. So, in this model, effectively, three body interactions are forbidden: if $i$ interacts with $j$, and $i$ interacts also with $k$, then $k$ and $j$ do not interact with each other. Thus, we can immediately notice that this model is mean-field in the sense that the network of interactions is tree-like, i.e., there are no \emph{loops}. Indeed, another possible mean field model for the glass state, the Mari-Kurchan-Krzakala model \cite{MKK}, embraces this philosophy explicitly by imposing a tree-like interaction network from the very start:
$$
H_{MKK} = \sum_{i,j}G_{ij}V(\bx_{i}-\bx_{j}),
$$
where $G_{ij}$ is the adjacency matrix of the underlying tree graph. Besides this, the disappearance of loops is also the mechanism that gives high-$d$ fluids their mean field nature. In that case, three-body interactions are made improbable (impossible for $d\to\infty$) by the high dimensionality itself.

\subsection{Partition function and entropy functional}
The canonical partition function for the model is 
$$
Z(\bA) = \int d^{N}\bx\ \exp\left(-\sum_{i<j}V(\bx_{i}-\bx_{j}-\bA_{ij})\right).
$$
Notice the absence of the factor $\frac{1}{N!}$, as particles in this model are not truly indistinguishable for a given realization of the random shifts. As usual, we want to compute the entropy of the liquid
$$
S(\varphi) = \dav{\log Z(\bA)}.
$$
In the liquid phase, the average over the shifts can be treated at an annealed level\cite{pedestrians}, that is, we can assume them to fluctuate on the same timescale as the system's microscopic configuration, thus giving
$$
S(\varphi) = \log \dav{Z(\bA)}.
$$
This way, our problem reduces to the computation of the annealed partition function
\begin{equation}
\begin{split}
\dav{Z(\bA)} =& \frac{1}{N!}\int\prod_{l<m}dP(\bA_{lm}) \int d^{N}\bx\ \exp\left(-\sum_{i<j}V(\bx_{i}-\bx_{j}-\bA_{ij})\right)\\
=&\int d^{N}\bx\ \prod_{i<j}\left(1+\dav{f}(\bx_{i}-\bx_{j})\right),
\label{MKgran}
\end{split}
\end{equation}
where $\dav{f}$ is the annealed Mayer function, and we have reinserted the factor $\frac{1}{N!}$ for convenience. Let us compute $\dav{f}$
\begin{equation}
\begin{split}
\dav{f}(\bx-\by) =& \int d\bA P(\bA) [e^{-V(\bx-\by-\bA)}-1]\\
=&-\int d\bA P(\bA) \theta(D-|\bx-\by-\bA|) \\
=&-\frac{1}{V} \int d\bA\ \theta(D-|\bx-\by-\bA|).
\end{split}
\end{equation}
We can see that the integral corresponds to the volume of a sphere of radius $D$ at position $\bx-\by$, so that
$$
\dav{f} = -\frac{v_{d}(D)}{V}.
$$

In order to compute the entropy $S$, we now turn to expansion \refeq{F}. As we anticipated, of the diagrams that are in the \refeq{F} only the first one survives, along the ideal gas term. So the entropy would be (notice the extra $(N\log N-N) \simeq \log N!$ factor, due to the non-indistinguishability of the particles)
\begin{equation}
S(\varphi) = -\int d\bx\ \rho(\bx)[\log\rho(\bx)-1] + \frac{1}{2}\int d\bx d\by\ \rho(\bx)\rho(\by)\dav{f}(\bx-\by) + N\log N -N.  
\label{FPentropy}
\end{equation}
To prove this, one can use diagrammatic theory by starting from the grancanonical expansion in \refeq{Omega}, which is actually the way it is done in \cite{marikurchan}. Here we will show it in a more straightforward manner, using the saddle point method.

We start from the partition function \refeq{MKgran}
$$
\int d^{N}\bx\ \prod_{i<j}\left(1+\dav{f}(\bx_{i}-\bx_{j})\right),
$$
let us define the density field
$$
\rho(\bx) = \sum_{i=1}^{N}\de(\bx-\bx_{i}).
$$
We now introduce a factor 1 in the partition function in the form of the field integral\footnote{This procedure is similar to the coulomb-gas method for large random matrices \cite{coulombgas}.}
$$
\int \de\rho(\bx)\ \de\left(\rho(\bx) - \sum_{i=1}^{N}\delta(\bx-\bx_{i})\right),
$$
and we get
\begin{equation}
\begin{split}
\dav{Z(\bA)} = &\int \de\rho(\bx)\int d^{N}\bx\ \de\left(\rho(\bx) - \sum_{i=1}^{N}\delta(\bx-\bx_{i})\right)\\
&\times\exp\left(\frac{1}{2}\int d\bx d\by\rho(\bx)\rho(\by)\log[1+\dav{f}(\bx-\by)]\right).
\end{split}
\end{equation}

Let us focus on the first line, i.e., on the entropic term. As usual we can express the delta function as a functional Fourier integral. We get
\begin{equation}
\begin{split}
&\int \de\hr(\bx)\ \de\rho(\bx)\ d^{N}\bx \exp\left[i\int d\bx \rho(\bx)\hr(\bx) - i\sum_{i=1}^{N}\int d\bx\hr(\bx)\de(\bx-\bx_{i})\right]\\
=&\int \de\hr(\bx)\ \de\rho(\bx)\ d^{N}\bx \exp\left[i\int d\bx \rho(\bx)\hr(\bx) - i\sum_{i=1}^{N}\hr(\bx_{i})\right].
\end{split}
\end{equation}
We can now put it all together, getting  
\begin{equation}
\dav{Z(\bA)} = \int \de\hr(\bx)\ \de\rho(\bx)\ e^{G[\rho(\bx),\hr(\bx)]},
\end{equation}
where the field action $G$ is 
$$
G[\rho(\bx),\hr(\bx)] = i\int d\bx\ \rho(\bx)\hr(\bx) + N\log\int d\bx\ e^{-i\hr(\bx)} + \frac{1}{2}\int d\bx d\by\rho(\bx)\rho(\by)\log[1+\dav{f}(\bx-\by)].
$$
Now, noticing that the Mayer function
$$
\dav{f}(\bx-\by) = -\frac{v_{d}(D)}{V} 
$$
goes to zero in the thermodynamic limit, we can expand the logarithm, getting 
$$
G[\rho(\bx),\hr(\bx)] = i\int d\bx\ \rho(\bx)\hr(\bx) + N\log\int d\bx e^{-i\hr(\bx)} + \frac{1}{2}\int d\bx d\by\rho(\bx)\rho(\by)\dav{f}(\bx-\by).
$$
To evaluate the functional integral in the thermodynamic limit, we can use the saddle point method with respect to the fields $\rho$ and $\hr$. The saddle point equations read
\begin{eqnn}
\rho(\bx) &=& N\frac{e^{-i\hr(\bx)}}{\int d\bx\ e^{-i\hr(\bx)}},\\
\hr(\bx) &=& i\int d\by\ \rho(\by)\dav{f}(\bx-\by).
\end{eqnn}
We can eliminate $\hr$ using the first one, and in the end we get
$$
\log\dav{Z(\bA)} = -\int d\bx\ \rho(\bx)\log\rho(\bx) + \frac{1}{2}\int d\bx d\by\ \rho(\bx)\rho(\by)\dav{f}(\bx-\by) + N\log N,
$$ 
which is equal to the \refeq{FPentropy}. The density profile $\rho(\bx)$ must in turn satisfy the equation
$$
\log\rho(\bx) = \int d\by\rho(\by) \dav{f}(\bx-\by),
$$
which can be derived by functional differentiation of the $\dav{Z(\bA)}$ or by elimination of $\hr(\bx)$ in the second of the saddle point equations. This integral equation corresponds to the minimization condition \refeq{eqrho} for the equilibrium density profile.

\subsection{Replica formalism}
We now wish to apply replica formalism to the study of the dynamical glass transition in the MK model. We remind that our aim is to compute the entropy $\Sc(m,\varphi,A)$ of $m$ coupled replicas of the model, and then look for any non-trivial solutions of the equation
$$
\dtot{\Sc}{A} = 0.
$$

In the glass phase, the average over the couplings must be done at a quenched level\cite{pedestrians} (the shifts are now frozen while the system evolves), i.e.
$$
\Sc(m,\varphi,\e) = \dav{\log Z_m},
$$
which is pretty hard to compute. We circumvent this difficulty using the \emph{replica trick}
$$
\dav{\log Z_m} = \lim_{n\to 0} \log\dav{Z_m^n} = \lim_{n\to 0} \frac{\dav{Z^n_m}-1}{n},
$$
that is, we consider $n$ uncoupled replicas of the system of $m$ weakly coupled replicas ($nm$ replicas in total). Thus, the annealed partition function would be
\begin{equation}
\begin{split}
\dav{Z_m^n}(\epsilon) = &\int \prod_{l<m}dP(A_{lm})\int \prod_{c=1}^n d^N\ox\\ 
&\times\exp\left(-\sum_{a=1}^{nm}\sum_{i<j}V_{HS}(\bx^a_i-\bx^a_j-\bA_{ij}) - \sum_{c=1}^n\sum_{a<b}^{a,b \in \{m_c\}}\sum_{i=1}^N\frac{\epsilon}{m}(\bx^a_i-\bx^b_i)^2\right) 
\end{split}
\end{equation}
where with $m_c$ we denote the $c$-th (of $n$ total) block of $m$ replicas. As we can see, the $nm$ replicas are grouped in blocks of $m$ each, and interact only within the same block via the harmonic potential.

As we anticipated, we now switch to the density-functional form of the entropy. For the $nm$-replicated system we have
\begin{equation}
\log\dav{Z^n_m} = -\int d\ux\ \rho(\ux)\log\rho(\ux) + \frac{1}{2}\int d\ux d\uy\ \rho(\ux)\rho(\uy)\dav{f}_r(\ux-\uy) + N\log N
\label{Snm}
\end{equation}
where we have defined
\begin{eqnn}
\ux &\equiv& (\ox_1,\ox_2,\dots,\ox_n) = (\bx_1,\dots,\bx_m,\bx_{m+1},\dots,\bx_{mn}),\\
\dav{f}_r(\ux-\uy) &\equiv& (-1)^{mn}\int d\bA P(\bA)\ \prod_{a=1}^{nm} \theta(D-|\bx_a-\by_a-\bA|).
\end{eqnn}
To obtain this form of the entropy, one starts from the partition function for the $nm$ replicated system and then generalizes simply the steps done in the preceding section.

Now, we must generalize the ansatz \refeq{Gaussian} to the case of $nm$ replicas coupled in blocks of $m$. This is indeed remarkably easy: as replicas are coupled only within blocks, then the ansatz for $\rho(\ux)$ will just be a product of $n$ Gaussian forms like the \refeq{Gaussian},
$$
\rho(\ux) \propto \prod_{c=1}^n\rho(\ox^c),
$$
and more precisely
\begin{equation}
\rho(\ux) = \frac{N}{V^n} \prod_{c=1}^n \int d\bX^c \frac{1}{(2\pi A)^{md/2}}\exp\left(-\sum_{a=m(c-1)+1}^{mc}\frac{(\bx^a-\bX^c)^2}{2A}\right).
\end{equation}

Now, to compute the entropy, we must plug this form of the density field into the \refeq{Snm}, compute it as a function of $n$, and then take the limit $n\to0$,
$$
\Sc(m,\varphi,A) = \dav{\log Z_m} = \lim_{n\to 0} \log\dav{Z^n_m}.
$$
The calculations are quite cumbersome, but in the end it can be shown that, once the $n\to 0$ limit has been taken, the result for $\Sc$ is 
\begin{equation}
\Sc(m,\varphi,A) = -\int d\ox\ \rho(\ox)[\log\rho(\ox) -1] + \frac{1}{2}\int d\ox d\oy\ \rho(\ox)\rho(\oy)f(\ox-\oy) + N\log N-N,
\label{Sm}
\end{equation}
where
$$
f(\ox-\oy) \equiv \prod_{a=1}^m \chi(\bx_a-\by_a) -1 ,
$$
and $\rho(\ox)$ is given by the \refeq{Gaussian}. Thus we have discovered that, apart from the $N\log N -N$ additive term, the entropy for the replicated MK model corresponds to the first two terms of expansion \refeq{F} for the replicated ordinary HS liquid: the advantage is that for the HS liquid this would be a (very crude at best) approximation, while for the MK model it is exact.

\subsection{The dynamical transition and the phase diagram}
We now compute, finally, expression \refeq{Sm} using the ansatz \refeq{Gaussian}. Let us start from the ideal-gas term
\begin{equation}
\begin{split}
\frac{1}{N}\int d\ox\ \rho(\ox)[1-\log\rho(\ox)] =& 1-\log\rho+(m-1)\frac{d}{2}\log (2\pi A) -\frac{d}{2}\log m\\ 
&+\frac{1}{N}\int d\ox\ \rho(\ox)\left[\frac{1}{2mA}\sum_{a<b}(\bx^a-\bx^b)^2\right].
\end{split}
\end{equation}
We can rewrite the integral in the second line as
$$
\frac{\rho}{N}\int d\bX\int \prod_{a=1}^m d\bx \gamma_A(\bx^a-\bX) \frac{1}{2mA}\left[(m-1)\sum_{a=1}^m (\bx^a)^2 -\sum_{a\neq b}\bx_a\cdot\bx_b\right],
$$
which can be evaluated easily, giving
$$
\frac{d(m-1)}{2}.
$$
Thus, the total ideal-gas entropy per particle, taking into account the ($N \log N -N$) additive factor\footnote{Since we have to take the derivative with respect to $A$, the $\log N$ factor is effectively harmless for what concerns the dynamical transition. However, this is not the case for the ideal glass transition, since to study it we must look for solutions of the equation $\Sigma(\varphi)=0$, and if the $\log N$ factor is considered part of the complexity, the transition disappears in the thermodynamic limit. For this reason, the MK model is not suitable for the study of the ideal glass transition. See \cite{charbonneaujin14} for more details.}, is
\begin{equation}
\begin{split}
S_{id}(m,A) =& -\log\rho+(m-1)\frac{d}{2}\log (2\pi A) -\frac{d}{2}\log m + d\frac{m-1}{2} + \log N\\
=& -\log\rho + \log N + S_{harm}(m,A),
\end{split}
\end{equation}
where 
$$
S_{harm}(m,A) \equiv (m-1)\frac{d}{2}\log (2\pi A) -\frac{d}{2}\log m + d\frac{m-1}{2}.
$$
We now turn to the interaction term
\begin{equation}
\frac{1}{2N}\int d\ox d\oy\ \rho(\ox)\rho(\oy)f(\ox-\oy).
\label{int}
\end{equation}
First, we rewrite the Mayer function in the following way
$$
\prod_{a=1}^m \chi(\bx_a-\by_a) -1 = \chi(\bx_1-\by_1)\left(\prod_{a=2}^m \chi(\bx_a-\by_a) -1 \right) + \chi(\bx_1-\by_1) - 1,
$$
and the integral \refeq{int} becomes
\begin{equation}
\begin{split}
&\frac{1}{2N}\int d\ox d\oy\ \rho(\ox)\rho(\oy)\chi(\bx_1-\by_1)\left(\prod_{a=2}^m \chi(\bx_a-\by_a) -1 \right)\\ 
+ &\frac{1}{2N}\int d\ox d\oy\ \rho(\ox)\rho(\oy)[\chi(\bx_1-\by_1) - 1].
\end{split}
\label{int2}
\end{equation}
The second integral can be computed trivially and the result is 
$$
\frac{\rho^2 V}{2N}\int d\bx [\chi(\bx)-1] = -\frac{\rho}{2}v_d(D) = -\frac{\rho}{2}2^dv_d(D/2) = -2^{d-1}\varphi.
$$
We now focus on the first one, which is (unsurprisingly) the most difficult. Indeed, it represents the ``correction'' to the interaction term which is due to the presence of the replicas.

We define the function
\begin{equation}
\begin{split}
Q(\bx-\by) \equiv &\int d\bx^2\cdots d\bx^m d\by^2\cdots d\by^m d\bX d\bY\\ 
&\gamma_{A}(\bx-\bY)\gamma_{A}(\by-\bY)\prod_{a=2}^m \gamma_{A}(\bx_a-\bX)\gamma_{A}(\by_a-\bY)[\prod_{a=2}^m \chi(\bx_a-\by_a) -1],
\end{split}
\label{Q}
\end{equation}
which would be the first effective interaction potential of the small-cage expansion \cite{parisimezardprl99}. This definition allows us to write the first line of the \refeq{int2} as 
\begin{equation}
\begin{split}
&\frac{1}{2N}\int d\ox d\oy\ \rho(\ox)\rho(\oy)\chi(\bx_1-\by_1)\left(\prod_{a=2}^m \chi(\bx_a-\by_a) -1 \right)\\ 
= &\ \frac{\rho^2}{2N}\int d\bx_1 d\by_1 \chi(\bx_1-\by_1)Q(\bx_1 - \by_1)\\
= &\ \frac{\rho v_d(D)}{2}\left(\frac{1}{v_d(D)}\int d\bx\chi(\bx)Q(\bx)\right)\\
=&\ 2^{d-1}\varphi G_m(A),
\end{split}
\end{equation}
with the definition
$$
G_m(A) = \frac{1}{v_d(D)}\int d\bx\ \chi(\bx)Q(\bx).
$$

We now need to compute $Q(\bx)$. First, we notice that since the $\gamma_A$ are normalized, we can rewrite the \refeq{Q} as
$$
\int d\bX d\bY\ \gamma_{A}(\bx-\bY)\gamma_{A}(\by-\bY)\left[ \left(\int d\bx d\by \gamma_{A}(\bx-\bX)\gamma_{A}(\by-\bY)\chi(\bx-\by)\right)^{m-1} -1\right],
$$
which prompts us to define another function
\begin{equation}
\begin{split}
q_A(\bX-\bY) =&\ \int d\bx d\by \gamma_{A}(\bx-\bX)\gamma_{A}(\by-\bY)\chi(\bx-\by)\\
=&\ \int d\xi d\eta\ \gamma_{A}(\xi)\gamma_{A}(\eta)\chi(\bX + \xi-\bY-\eta)\\
=&\ \int d\br'\ \gamma_{2A}(\br')\chi(\bX-\bY - \br').
\end{split}
\end{equation}
The last step was accomplished by changing coordinates from $(\xi,\eta)$ to $(\xi+\eta,\xi-\eta)$ and then evaluating the Gaussian integral on $\xi+\eta$.\\
Now, for the $Q$ we have
$$
Q(\bx-\by) = \int d\bX d\bY\ \gamma_{A}(\bx-\bY)\gamma_{A}(\by-\bY)\left[q_A(\bX-\bY)^{m-1} -1\right],
$$
which can be manipulated in the same way as the $q_A$, getting
$$
Q(\bx-\by) = \int d\br'\ \gamma_{2A}(\br')[q_A(\bx-\by-\br')^{m-1} -1].
$$
Plugging this in the $G_m(A)$, we then have
\begin{eqnn}
G_m(A) &=& \frac{1}{v_d(D)}\int d\br\chi(\br)Q(\br)\\
&=& \frac{1}{v_d(D)}\int d\br[q_A(\br)^{m} - \chi(\br)].
\end{eqnn}

At this point, we can rewrite the whole entropy
\begin{equation}
\frac{\Sc(m,\varphi,A)}{N} = -\log\rho + \log N + S_{harm}(m,A) - 2^{d-1}\varphi[1-G_m(A)].
\end{equation}
Now, to compute the equilibrium value of $A$ (remember equation \refeq{gibbs}) we take its derivative with respect to $A$ and equate it to zero:
$$
\dtot{\Sc}{A} = 0,
$$
which gives the equation
\begin{equation}
\frac{d}{2^d \varphi} = \frac{A}{m-1}\dpart{G_m(A)}{A} \equiv F_m(A),
\end{equation}
where we have defined the function
$$
F_m(A) = \frac{mA}{m-1}\frac{1}{v_d(D)} \int d\br\ q_A(\br)^{m-1}\dpart{q_A(\br)}{A}.
$$
Since we are interested in the dynamical transition for the real, non-replicated liquid, we must send $m$ to one (one replica only). Taking (with some caution) the limit of the $F_m(A)$ for $m\to 1$, we get
\begin{equation}
F_1(A) = -\frac{A}{v_d(D)}\int d\br \log[q_A(\br)]\dpart{q_A(\br)}{A}.
\label{F1}
\end{equation}
So, all that is left is computing the function $F_1(A)$, which is defined in terms of $q_A(\br)$.\\
We remind the definition of $q_A(\br)$:
$$
q_A(\br) \equiv \int d\br'\ \gamma_{2A}(\br')\chi(\br-\br').
$$
We observe that $q_A$ is the convolution in $d$ dimensions of a Gaussian with a theta-like function. This would prompt us to compute it by Fourier-transforming the two functions, but this procedure would lead to problems: since the cage radius $A$ is small, then the $\gamma_{2A}$ has a small variance, which implies that its Fourier transform will be long-ranged. The same would apply to $\chi$ as its range is equal to the sphere diameter, which is more or less of the same order of magnitude as $A$.\\
Thus, rather than Fourier-transforming back the product of two long-ranged functions, it is more convenient to evaluate the convolution by using $d$-dimensional bipolar coordinates, as it is done in \cite{parisizamponi}. This allows us to write $q_A(r)$ as a one-dimensional integral where $d$ appears as a parameter. As the calculations are quite long and tedious, we skip directly to the final result
\begin{equation}
q_A(r) = \int_D^\infty du\ \left(\frac{u}{r}\right)^{(d-1)/2}\frac{e^{-(r-u)^2/4A}}{\sqrt{4\pi A}}\left[e^{-ru/2A}\sqrt{\frac{\pi r u}{A}}I_{(d-2)/2}\left(\frac{ru}{2A}\right)\right],
\end{equation}
where $I_{i}(x)$ is $i$-th order modified Bessel function of the first kind. This integral can be evaluated explicitly only under certain conditions, depending on the value of $d$.\\ 
More precisely, if $d$ is odd, the function $I_{(d-2)/2}$ has an analytic expression in terms of hyperbolic functions and polynomials, and the $q_{A}(r)$ can be computed analytically, although the resulting expression becomes more and more cumbersome with higher $d$. For $d=3$ for example, $q_{A}(r)$ takes the form
\begin{equation}
\begin{split}
q_{A}(r) =& \frac{1}{r\sqrt{4\pi A}} \int_{D}^{\infty}du\ u[e^{-(r-u)^{2}/4A}-e^{-(r+u)^{2}/4A}]\\
=& \frac{1}{2}\left[\erf\left(\frac{r-D}{\sqrt{4A}}\right)-\erf\left(\frac{r+D}{\sqrt{4A}}\right) + \frac{2}{r}\sqrt{\frac{A}{\pi}}(e^{-(r-D)^{2}/4A} - e^{-(r+D)^{2}/4A}) +2\right].
\end{split}
\end{equation}
In the limit of high dimension $d\to\infty$, which is the mean-field limit for an ordinary HS liquid, one can extract the asymptotics by using the saddle point method on the integral representation of Bessel functions. In this case, the correct scaling form for $A$ can be found to be 
$$
A = \frac{D^{2}\hat{A}}{d^{2}},
$$
so everything must be expressed as a function of the scaling variable $\hat{A}$; for further details see \cite{parisizamponi,parisikurchan}.\\
For $d$ even, unfortunately, the Bessel function does not have an analytic expression and the $q_{A}(r)$ must then be computed numerically.

\begin{figure}[htb!]
\begin{center}
\includegraphics[width=0.5\textwidth]{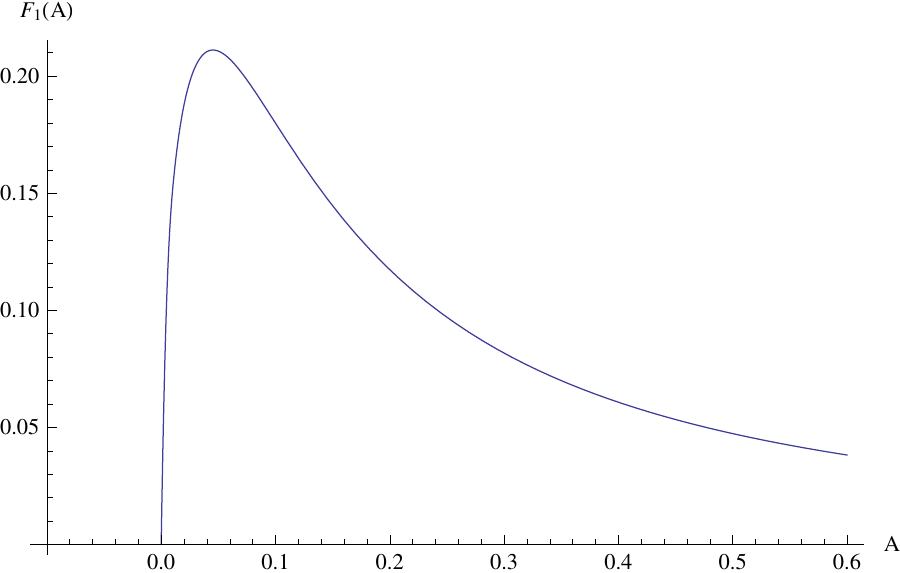}\nolinebreak[4]
\includegraphics[width=0.5\textwidth]{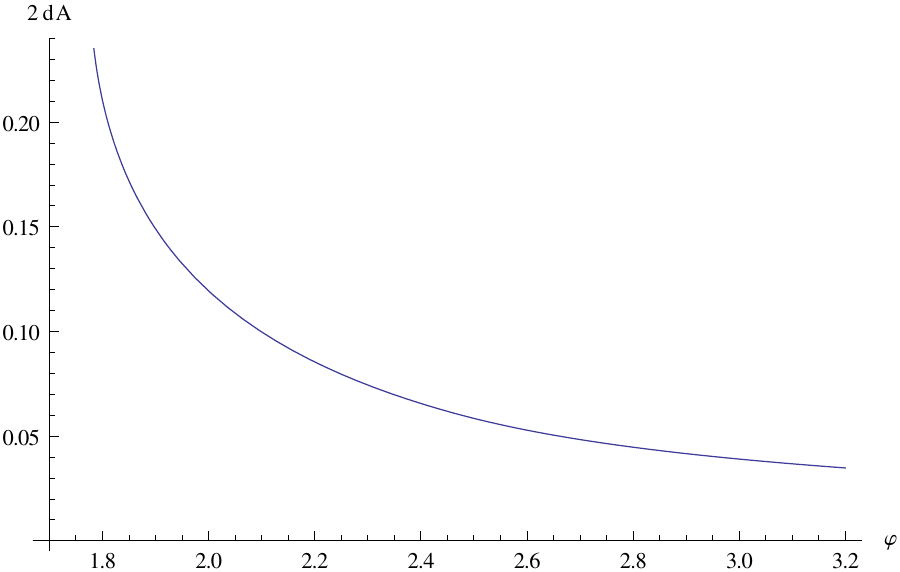}
\caption{Left: the function $F_{1}(A)$ for $d=3$. Right: The cage radius $A$ as a function of the packing fraction $\varphi$, for $\varphi>\varphi_{d}$, $d=3$.}
\label{F1fig}
\end{center}
\end{figure}

\begin{figure}[htb!]
\begin{center}
\includegraphics[width=0.5\textwidth]{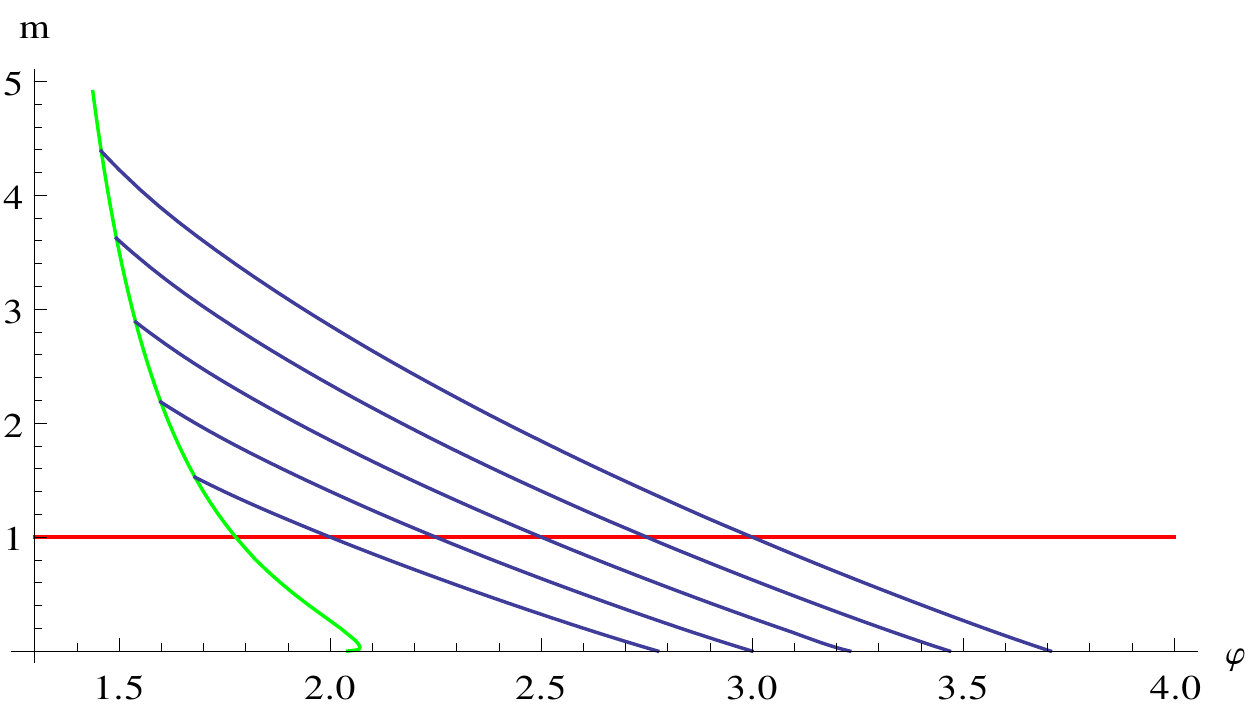}
\caption{The $(\varphi,m)$ phase diagram for the MK model in three dimensions. The green line is the dynamical transition line. The blue lines are isocomplexity lines, i.e. the contours of the complexity as a function of $\varphi$ and $m$. Reprinted from \cite{corrado}.}
\label{emmephi}
\end{center}
\end{figure}

Once $q_{A}(r)$ is known, one must compute the \refeq{F1}. In this case the integral must be always evaluated numerically, so at the moment we don't have an analytic expression for $F_{1}(A)$; this is also why the study of the even $d$ case is quite error prone, since one must then perform a numerical integration using the result of a numerical integration as an input.\\
Nevertheless, $F_{1}(A)$ can be computed easily in dimension $3$ using Mathematica (we fix $D=1$), and its plot is shown in the left panel of figure \ref{F1fig}.
We recall the equation for the cage radius
$$
\frac{d}{2^d \varphi} \equiv F_1(A).
$$
As we can see, for low $\varphi$ the equation has no solution, since $F_{1}(A)$ is bounded. As soon as the constant on the left side touches the maximum of $F_{1}(A)$, two nontrivial solutions for $A$ appear, of which the physical one is the one which decreases with increasing $\varphi$. The plot of $A$ vs. $\varphi$ is shown in the right panel figure \ref{F1fig}.\\    
The dynamical transition density $\varphi_{d}$ can be computed as 
$$
\varphi_{d} = \frac{d}{2^{d}\textrm{max}_{A}\{F_{1}(A)\}},
$$ 
ad we get 
$$
\varphi_{d} \simeq 1.7764,
$$
which is a reasonable value \cite{marikurchan,charbonneaujin14}. In addition, using the generic $F_{m}(A)$, we can compute $\varphi_{d}$ as a function of $m$ and thus draw the dynamical transition line in the $(\varphi,m)$ plane. We show the $(\varphi,m)$ phase diagram from the model in figure \ref{emmephi}.

\section{Conclusions and perspectives}
We presented here a brief review of the replica method in its application to the structural glass transition problem. We started from an intuitive picture of how the method works, and from there we have gone all the way to the computation of the phase diagram for an exemplary model. Of course, there are many more results that can be obtained using replica theory. Indeed, the phenomenology of the MK model is extremely rich, despite its simplicity, and shows a number of non-trivial features such as non-perfect caging of particles due to hopping, and violation of the Stokes-Einstein relation for viscosity and diffusion \cite{charbonneaujin14}. In addition to this, the presence of the random shifts makes possible the use of a procedure, the \emph{planting}\cite{krzakalazdeborovaplanting} method, which allows us to obtain equilibrated configurations for densities deep in the glass phase. This means that we can circumvent the problem of the extremely large equilibration times needed to thermalize a glass former 
near the glass transition, opening 
the 
door to extensive numerical studies of the high density, glassy regime\cite{corrado}; in particular, we can simulate the actual process of glass formation by a slow annealing to a planting temperature $T_g<T_d$, where the glass forms, and then model the behavior of the so obtained glass as it is heated or cooled via rapid temperature variations. This process is the one that is actually used to make glasses in the real world and so the study of this regime is extremely important for all practical and experimental purposes.\\
Since in this regime the glass former is trapped inside a single metastable state, the theoretical focus must shift to the in-state free entropy $s$ which actually governs the physical properties of the liquid when it is far from equilibration. The replica method can be used in this case as well, along with the \emph{isocomplexity} \cite{montanariricciisocomplexity} assumption, to compute the properties of the glass, as it is shown in \cite{corrado}, where the comparison to numerics and the link to actual experiments on glass formers is discussed. A more refined formalism, which relies on the Franz-Parisi potential \cite{franzparisi} has been applied in \cite{corrado2} to the infinite-dimensional liquid studied in \cite{parisikurchan,zamponiurbani,zamponifull}, where the issue of the response of the glass to an external drive has also been addressed.\\
In general, we can say that since the seminal paper \cite{ParisiMezard}, down to the more recent results of the series \cite{parisikurchan,zamponiurbani,fullRSB,zamponifull}, the replica method has thoroughly proven its worth for what concerns the study of glass forming systems in the mean-field limit, where metastable glassy states have an infinite lifetime and ergodicity breaking is hard. However, one must not forget that the presence and even the nature of metastable states is still a debated issue (see the conclusions of \cite{zamponi} for details) for all cases in which the system does not remain trapped forever inside them and ergodicity breaking is not hard. Once this happens, the concept of ``state" could actually be not meaningful anymore. Unfortunately, this is what actually happens in nature, as the lifetime of metastable states is indeed very long, but finite even below the dynamical transition temperature. Thus, we need to consider realistic, out of mean field models for the RFOT program for the 
description 
of the glass transition to reach its goal. The applicability of the replica method in those situations will then be undoubtedly linked to the fate of metastable states once the system is brought out of mean field.

\section{Acknowledgements}
I wish to thank Hugo Jacquin for many useful suggestions, and Ferdinando Randisi for his precious comments on the manuscript. I also wish to thank Giorgio Parisi, Francesco Zamponi and Pierfrancesco Urbani for their support and advice.

\cleardoublepage

\bibliographystyle{amsplain}
\bibliography{bibliografia.bib}

\end{document}